\documentclass[11pt, a4paper]{statsoc}

\usepackage{epsfig}
\usepackage{verbatim}
\usepackage{graphicx}
\usepackage[authoryear]{natbib}
\usepackage{amsfonts}
\usepackage{amsmath}
\usepackage{amssymb}
\usepackage{graphics, color}
\usepackage{multicol}
\usepackage{multirow}
\usepackage{setspace}
\usepackage{float}
\usepackage{hyperref}

\newcommand{\bd}[1]{\ensuremath{\mbox{\boldmath $#1$}}}

\title[ Adaptive spatio-temporal smoothing for disease risk]{An adaptive spatio-temporal smoothing model for estimating trends and step changes in disease risk}
\author[Rushworth, Lee and Sarran]{Alastair Rushworth$^a$, Duncan Lee$^b$ and Christophe Sarran$^c$}
\address{$^a$Department of Mathematics and Statistics, University of Strathclyde, UK.}
\address{$^b$School of Mathematics and Statistics, University of Glasgow, UK.}
\address{$^c$UK Met Office, Exeter, UK.}
\email{alastair.rushworth@strath.ac.uk}
\begin{document}

\footnotetext[1]{\it{Email address for correspondence: alastair.rushworth@strath.ac.uk}}

\begin{abstract}
Statistical models used to estimate the spatio-temporal pattern in disease risk from areal unit data  represent the risk surface for each time period with known covariates and a set of spatially smooth random effects.  The latter act as a proxy for unmeasured spatial confounding, whose spatial structure is often characterised by a spatially smooth evolution between some pairs of adjacent areal units while other pairs exhibit large step changes. This spatial heterogeneity is not consistent with existing global smoothing models, in which partial correlation exists between all pairs of adjacent spatial random effects. Therefore we propose a novel space-time disease model with an adaptive spatial smoothing specification that can identify step changes. The model is motivated by a new study of respiratory and circulatory disease risk across the set of Local Authorities in England, and is rigorously tested by simulation to assess its efficacy. Results from the England study show that the two diseases have similar spatial patterns in risk, and exhibit a number of common step changes in the unmeasured component of risk between neighbouring local authorities.

\end{abstract}

%%%%%%%%%%%%%%%%%%%%%%%%%%%%%%%%%%%%%%%%%%%%%%%%%%
%%%%%%%%%% INTRODUCTION 
%%%%%%%%%%%%%%%%%%%%%%%%%%%%%%%%%%%%%%%%%%%%%%%%%%

\section{Introduction}

% Disease mapping wombling importance in public health (not academia); health inequalties; disease maps
\noindent Disease risk exhibits spatio-temporal variation due to many factors, including changing levels of environmental exposures and risk-inducing behaviours such as smoking. Disease risk data are typically obtained as population level summaries for administrative geographical units, such as local authorities, and the spatial pattern in risk is presented via a choropleth map. Such maps enable public health professionals to quantify the spatial pattern in disease risk, allowing financial resources and public health interventions to be targeted  at high-risk areas. Disease maps are routinely published by health agencies worldwide, such as the cancer e-Atlas (\url{http://www.ncin.org.uk/cancer_information_tools/eatlas/}) by Public Health England and the weekly influenza maps (\url{http://www.cdc.gov/flu/weekly/usmap.htm}) produced by the Centres for Disease Control and Prevention in the USA. These maps also allow the scale of health inequalities between rich and poor communities and their underlying drivers to be quantified. For example, a 2014 report by the \cite{ons2014} estimates that UK average healthy life expectancy differs by nineteen years between communities with the highest and the lowest levels of deprivation.  Such large inequalities exacerbate socioeconomic divisions in society, and health costs may be increased due to higher disease prevalence in the most disadvantaged regions.\\

% why model; key questions to address; why not raw maps of rates
\noindent Disease maps presented by health agencies display raw disease rates, which do not allow inferential statements to be made such as calculating risk-exceedence probabilities (\citealp{richardson2004}), or evaluating the significance of temporal changes.  A range of statistical models have been developed for disease data, which represent the risk surface using known covariates and a set of random effects.  The latter act as a proxy for unmeasured spatial confounding, and are typically modelled by    a Gaussian Markov Random field (GMRF, \citealp{rue2005gaussian}) prior in a hierarchical framework. GMRF priors have been extended to incorporate spatio-temporal structure, and prominent examples include \citet{bernardinelli1995bayesian}, \citet{knorr2000bayesianmodelling}, \citet{macnab2001autoregressive} and \citet{ugarte2010spatiotemporal}.\\

\noindent GMRF models assume the random effects are globally spatially smooth, in the sense that a single parameter governs the spatial autocorrelation in disease risk between all pairs of geographically adjacent units. In practice however, this residual or unexplained spatial structure is often characterised by a spatially smooth evolution between some pairs of adjacent units, while other pairs exhibit large step changes.  The identification of such step changes in the unexplained component of risk is known as \emph{Wombling}, following the seminal article by \citet{womble1951}, and can provide a number of epidemiological insights. Firstly, it allows the identification of the geographical extent of clusters of areal units that exhibit elevated unexplained risks, which enables health resources and public health interventions to be targeted appropriately. Secondly, it provides detailed insight into the spatial structure of the unmeasured confounding, allowing the identification of unknown etiological factors that contribute to disease risk. Furthermore, smoothing models that ignore local structure may result in oversmoothing in regions where strong disparities exist and undersmoothing elsewhere, leading to biased estimation of disease risk.  A range of spatially adaptive smoothing priors have been proposed to address these limitations for purely spatial data, including \citet{green2002hidden}, \citet{lu2007bayesian}, \citet{lawson2012bayesian}, \citet{lee2013locally},  \citet{wakefield2013bayesian} and \citet{lee2014bayesian}.\\

%  extensions to space time, deficiencies of current adaptive approaches
\noindent  Few spatially adaptive smoothing models have been developed for spatio-temporal disease data, with an exception being \citet{lee2014smmr} who propose an iterative fitting algorithm using integrated nested Laplace approximations.  While temporal replication is likely to improve the estimation in such highly complex models, the increased numbers of data points and parameters results in much increased computational complexity.  Therefore, the contribution of this paper is the development of a computationally feasible spatially-adaptive GMRF model for spatio-temporal disease data, which can be viewed as both an adaptive smoother and a model for the detection of step changes in unexplained risk. The model builds on the purely spatial approach of \citet{ma2010hierarchical}, and avoids making simplifying assumptions about the step-change structure as \citet{lee2014bayesian} do. Additionally, our model is freely available via the \texttt{R} package \texttt{CARBayesST} (\citealp{lee2015carbayesst}). The methodological development is motivated by a new study of respiratory and circulatory disease in England, UK, which according to the World Health Organisation (WHO) are two of the largest causes of death worldwide (\url{www.who.int/mediacentre/factsheets/fs310/en/}). This study is presented in Section \ref{sec:motivating}, while in Section \ref{sec:review} the literature on spatio-temporal disease mapping is reviewed. Section \ref{sec:model} proposes a new space-time GMRF model for adaptive smoothing,  which is comprehensively tested by simulation in Section \ref{sec:simulation}. In Section \ref{sec:application} the proposed model is applied to the motivating application, while the paper concludes in Section \ref{sec:discussion}.

\section{Motivating case study}
\label{sec:motivating}
\noindent Our methodological development is motivated by a new study of circulatory and respiratory disease risk in England between 2001 and 2010, which have International Classification of Disease tenth revision codes I00-I99 and J00-J99 respectively. Hospital admissions records from the Health and Social Care Information Centre (HSCIC) were analysed at the UK Met Office to provide yearly counts of emergency admissions by local and unitary authority (LUA). The resulting data $\{Y_{ij}\}$ are counts of hospital admissions for $i=1,\ldots,N$ ($N=323$) LUA in England in year $j=1,\ldots,T$ ($T=10$), and range between 6 -  1030  (circulatory) and  0 - 2485 (respiratory) respectively. The expected numbers of hospital admissions $\{E_{ij}\}$ were calculated for each year and LUA to adjust for their differing population sizes and demographic structures using indirect standardisation. Specifically, $E_{ij}=\sum_{r=1}^{q}N_{ijr}p_{r}$, where $N_{ijr}$ is the population size in LUA $i$, year $j$ and strata $r$ (e.g. males 0-5, etc), while $p_r$ is the England-wide risk of disease in strata $r$. The expected counts range between 144.8 - 564.1 (circulatory) and 115.2 - 675.9 (respiratory) respectively.\\

\begin{figure}
\centering
\begin{tabular}{cc}
 \includegraphics[clip, trim = 50mm 20mm 60mm 20mm, width = 6.4cm]{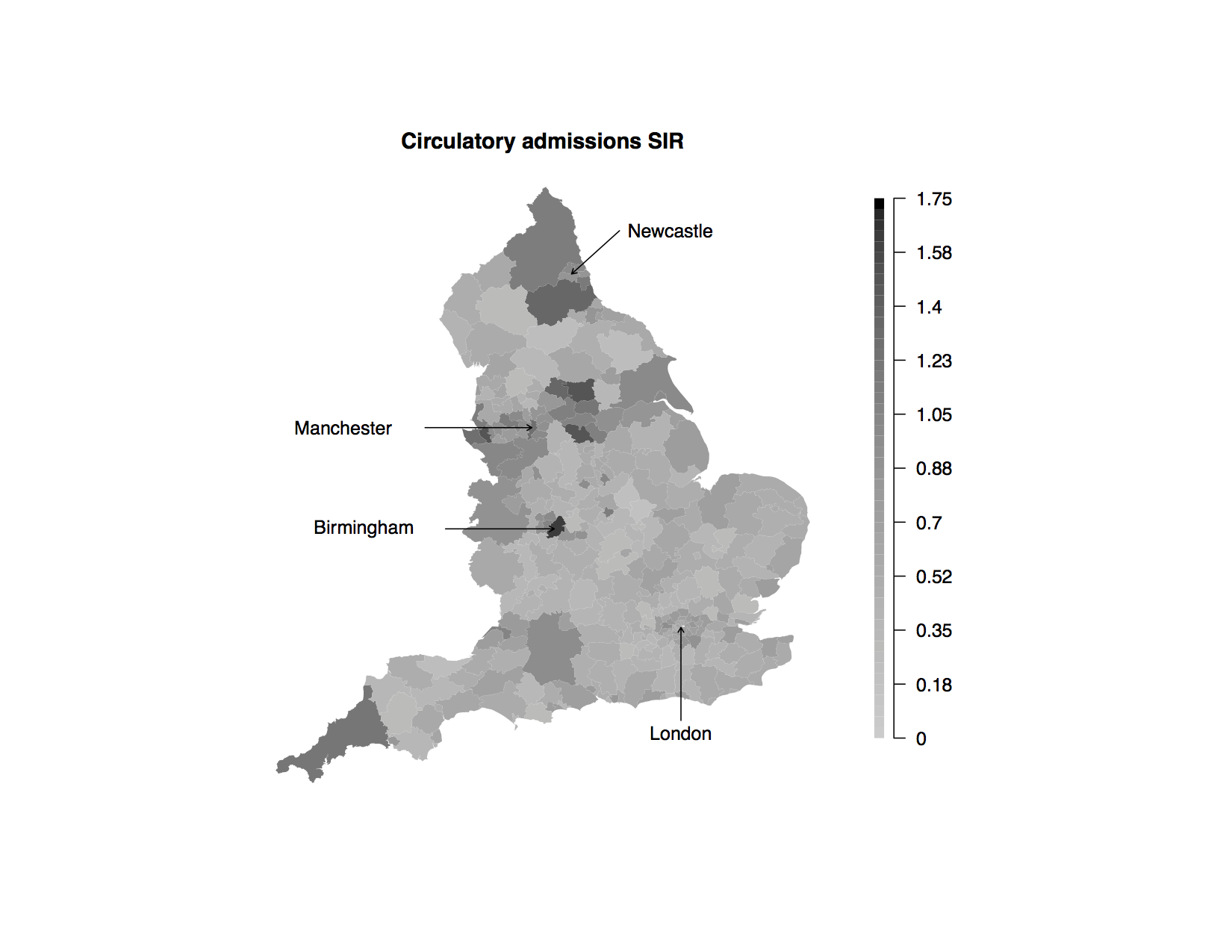}
& \includegraphics[clip, trim = 50mm 20mm 60mm 20mm, width = 6.4cm]{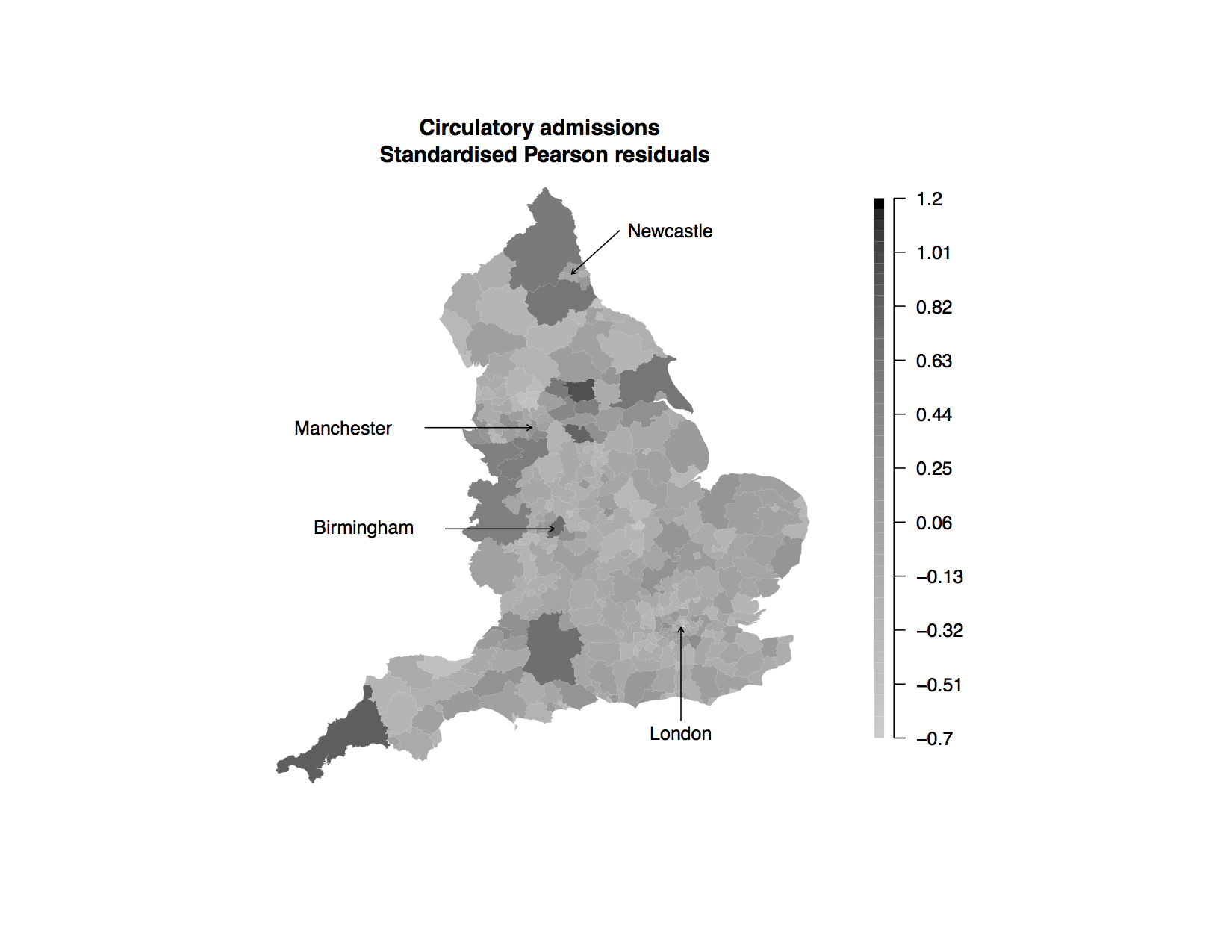} \\
 \includegraphics[clip, trim = 50mm 20mm 60mm 20mm, width = 6.4cm]{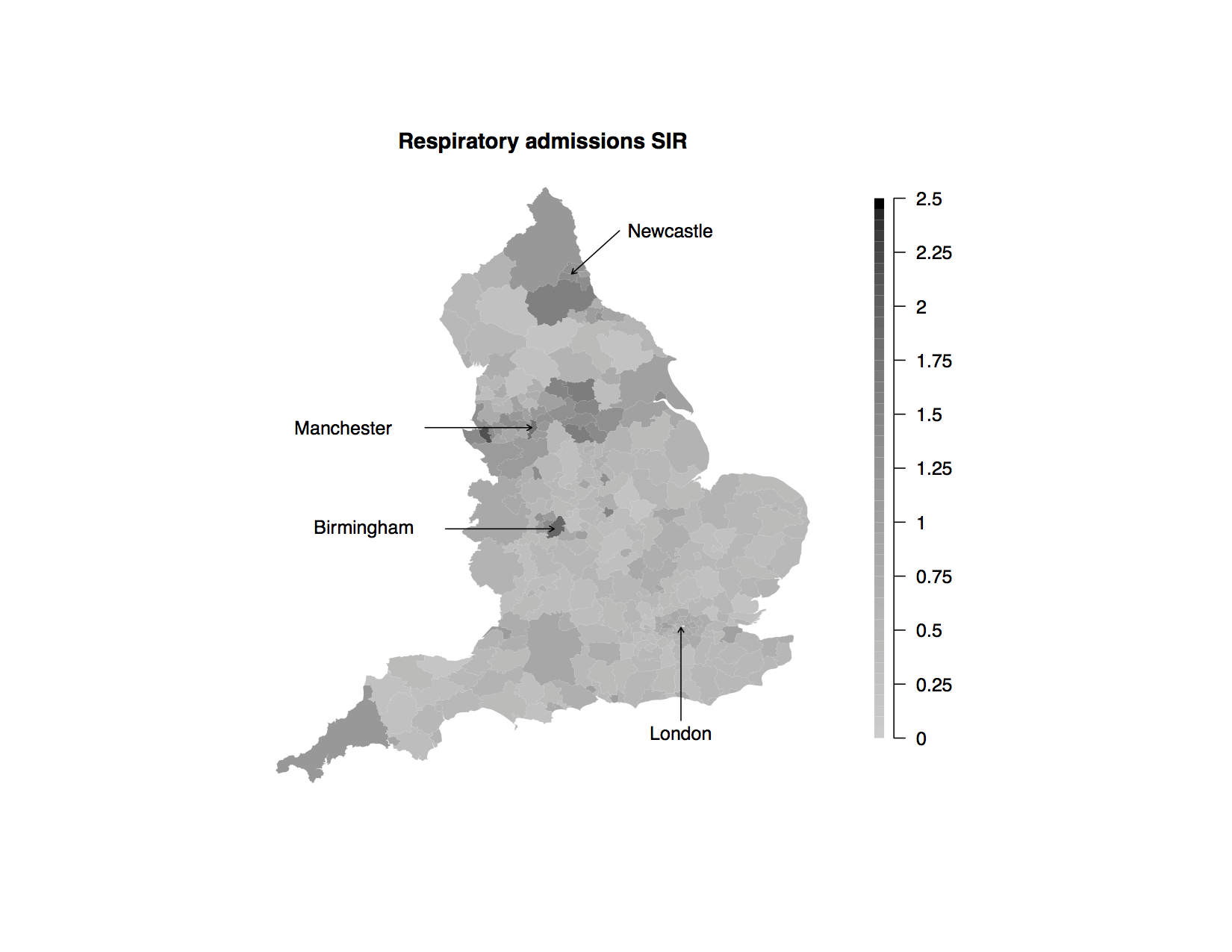}
& \includegraphics[clip, trim = 50mm 20mm 60mm 20mm, width = 6.4cm]{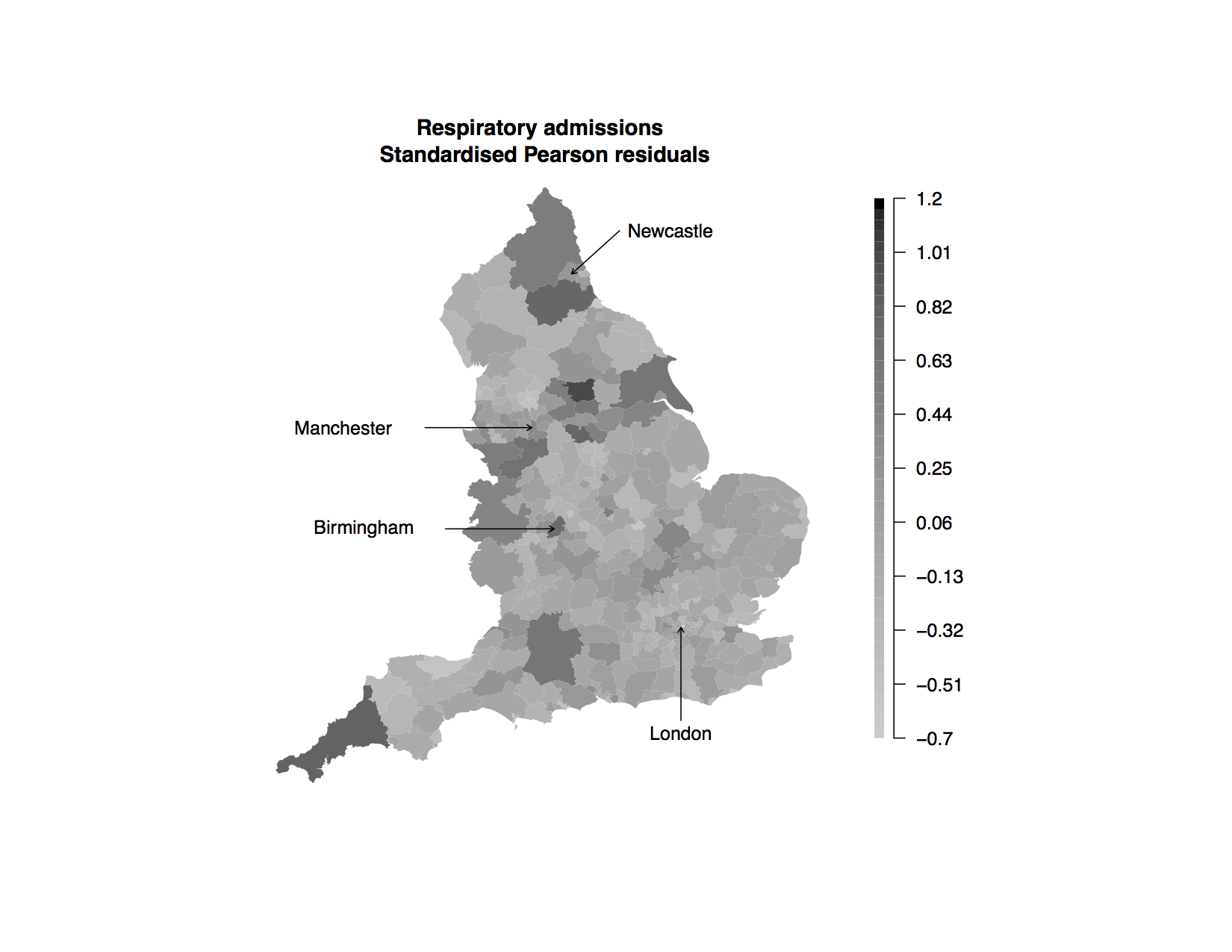} 
\end{tabular}
\caption{Left: Mean standardised incidence ratios (SIR) for circulatory (top left) and respiratory (bottom left) hospital admissions across English local authorities between 2001 and 2010.  Right: Mean standardised Pearson residuals from a Poisson log-linear model with JSA, urbanicity and PM$_{10}$ as covariates for circulatory (top right) and respiratory admissions (bottom right).  In all cases, the locations of four major English cities are indicated with arrows.}
\label{fig:SIR}
\end{figure}

The Standardised Incidence Ratio, SIR$_{ij}$=$Y_{ij}/E_{ij}$ is an exploratory (noisy) measure of disease risk, and a value of 1.2 corresponds to a 20$\%$ increased risk of disease relative to $E_{ij}$. The mean SIR over all years is displayed in the left column of Figure \ref{fig:SIR}, and similar spatial patterns are evident between the two diseases with a Pearson's correlation coefficient of 0.9356.  Within each map, the risk levels are spatially smooth across much of England, although a number of step changes are visible, including around the cities of Birmingham and Manchester.  There are a number of potential drivers of this spatial variation in disease risk, including socio-economic deprivation (poverty), air pollution and the differences between urban and rural areas. We measure poverty by the percentage of the working age population who are in receipt of Job Seekers Allowance (JSA) obtained from the HSCIC, while we obtain modelled particulate matter concentrations (PM$_{10}$) from the Department for Environment Food and Rural Affairs on a 1km grid which are then averaged to the LUA scale. Finally, the urban or rural nature of each LUA is measured by the proportion of middle layer super-output areas (MSOAs) classified as urban within each LUA.  The (mean over year) residuals from a simple Poisson log-linear model with these covariates are displayed in the right column of Figure \ref{fig:SIR}. The residual unexplained spatial variation in disease risk is autocorrelated for both diseases, with significant (at the 5$\%$ level) Moran's I statistics ranging between 0.204 and 0.328 across the years. However, Figure \ref{fig:SIR} also highlights that these unexplained spatial structures exhibit step changes, which are not compatible with a global spatial smoothing model. \\

The aims of modelling these data are twofold. First, we want to produce the best estimate of the spatio-temporal patterns in circulatory and respiratory disease risks, so that the extent of the health inequalities in these two diseases can be identified. Second,  we wish to estimate the locations of the step changes in the unexplained risk surface (Wombling), so that the geographical extent of clusters of excessively high unexplained risks regions can be identified and investigated for possible causes. To achieve these goals we propose an adaptive smoothing model in Section 4, but first present a review of the literature in Section 3.  In this paper we take a univariate approach and model each disease separately, as is standard practice in the space-time disease mapping literature. An alternative would be a bivariate spatio-temporal model, and the choice between a univariate or bivariate approach depends on the questions one wishes to address.  Here our interest is in developing one of the first spatially adaptive spatio-temporal smoothing models for disease risk, and comparing its efficacy to global smoothing alternatives. Such alternatives do not exist in a bivariate context, and thus we would be unable to assess the improvement our adaptive smoothing model provides in this setting. Furthermore, our interest is in estimating the presence, and similarity between diseases, of step changes in the unexplained component of disease risk, and a bivariate model would borrow strength across diseases and thus bias the results towards finding step changes common to both diseases.

\section{Spatio-temporal disease mapping}\label{sec:review}

\noindent The observed and expected disease counts for LUA $i$ and year $j$ are denoted by $(Y_{ij}, E_{ij})$ respectively, and the following Poisson log-linear model is commonly specified for these data:

\begin{eqnarray}
Y_{ij}|E_{ij}, R_{ij} &\sim & \textrm{Poisson}(E_{ij} R_{ij})~~~~i=1,\ldots,N,~~j=1,\ldots,T,\label{likelihood}\\
\ln (R_{ij}) & = & \mathbf{x}_{ij}^{\top}\bd{\beta} +\phi_{ij},\nonumber\\
 \beta_r & \sim & \mbox{N}(0, 10000) ~~~~r=1,\ldots,p\nonumber.
\end{eqnarray}

\noindent Disease risk is represented by $R_{ij}$, which is on the same scale as the SIR. The log-risk $\ln(R_{ij})$ is modelled by a vector of $p$ known covariates $\mathbf{x}_{ij} = (x_{ij1},\ldots,x_{ijp})$ with parameters $\bd{\beta}=(\beta_1,\ldots,\beta_p)$, and a spatio-temporal random effect $\phi_{ij}$.  GMRF priors are commonly used to induce spatial smoothness between the random effects, via a binary $N\times N$ adjacency matrix $\mathbf{W}$. Element $w_{ik} = 1$ if areas $i$ and $k$ share a common border (denoted $i\sim k$) and $w_{ik}=0$ otherwise (denoted $i\nsim k$), while $w_{ii}=0$ for all $i$. Numerous GMRF priors have been developed for purely spatial random effects $(\phi_{1},\ldots,\phi_{N})$, and the proposal by \citet{Leroux2000} has an attractive full conditional decomposition given by

\begin{eqnarray}
\phi_i|\bd{\phi}_{-i}, \rho, \tau^{2}, \mathbf{W} &\sim & \mbox{N}\left(\frac{\rho \sum_{k=1}^{N} w_{ik}\phi_k}{\rho\sum_{k=1}^{N}w_{ik} + 1 - \rho}, \frac{\tau^2}{\rho\sum_{k=1}^{N}w_{ik} + 1 - \rho}\right),\label{leroux}
\label{eqn:randeffect}
\end{eqnarray}

\noindent where $\bd{\phi}_{-i}=(\phi_{1},\ldots,\phi_{i-1},\phi_{i+1},\ldots,\phi_{N})$. The conditional expectation of $\phi_{i}$ is a weighted average of adjacent $\phi_k$ (as specified by $\mathbf{W}$), which induces smoothness across the surface. Spatial smoothing is controlled by $\rho \in [0,1]$, where $\rho=1$ corresponds to the intrinsic autoregressive model (\citealp{besag1974spatial}), while $\rho=0$ corresponds to independence. The resulting joint distribution is given by $(\phi_{1},\ldots,\phi_{N})\sim\mbox{N}(\mathbf{0}, \tau^2\mathbf{Q}(\rho,\mathbf{W})^{-1})$, where the precision matrix is given by $\mathbf{Q}(\rho,\mathbf{W})= \rho [\textrm{diag}(\mathbf{W}\textbf{1}) - \mathbf{W} ] + (1 - \rho) \mathbf{I}$, where $\textbf{1}$ is an $N\times 1$ vector of ones and $\mathbf{I}$ is the $N \times N$ identity matrix. \\

\noindent  Many spatio-temporal GMRF models have been developed in the disease mapping literature, with the first being \citet{bernardinelli1995bayesian} who model $R_{ij}$ with linear time-trends that have spatially varying slopes and intercepts.  In contrast, \citet{knorr2000bayesianmodelling} introduced a decomposition of $R_{ij}$ into spatial and temporal main effects and an interaction, with all terms being modelled by GMRF priors.  More recently, \citet{rushworth2014spatio} utilise the autoregressive decomposition 
\begin{equation}
f(\bd{\phi}_1, \ldots, \bd{\phi}_T)  =  f(\bd{\phi}_1) \prod_{j=2}^{T} f(\bd{\phi}_j|\bd{\phi}_{j-1}),\label{AR}
\end{equation}

where $\bd{\phi}_j=(\phi_{1j},\ldots,\phi_{Nj})$. They combine the decomposition (\ref{AR}) with the Leroux CAR prior for each $\bd{\phi}_j$, so that $\bd{\phi}_{1}$ is modelled by (\ref{leroux}), and $\bd{\phi}_j \sim\mbox{N}\left(\alpha \bd{\phi}_{j-1}, \hspace{0.1cm}\tau^2 \mathbf{Q}(\rho,\mathbf{W})^{-1} \right)$ for $j=2,\ldots,T$, where $\alpha \in [0,1]$ controls temporal autocorrelation. The global nature of the spatial autocorrelation induced by (\ref{leroux}) for purely spatial random effects $(\phi_{1},\ldots,\phi_{N})$ can be seen from their theoretical partial autocorrelations:
 
\begin{equation}
\textrm{Corr}[\phi_i, \phi_k|\bd{\phi}_{-ik},\rho,\mathbf{W}] = \frac{\rho w_{ik}}{\sqrt{(\rho\sum_{r=1}^N w_{ir} + 1-\rho)(\rho\sum_{s=1}^N w_{ks} + 1-\rho)}}.\label{partial}
\end{equation}
 
Under models (\ref{leroux}) and (\ref{AR}), random effects for all pairs of neighbouring areal units (for which $w_{ik}=1$) will be partially autocorrelated, and the strength of that partial autocorrelation will be controlled by $\rho$. Thus as $\rho$ will often be close to one (the spatial residual surfaces are autocorrelated as described in Section 2), a pair of adjacent areas exhibiting substantially different levels of unexplained risk will have those risks wrongly smoothed towards each other, masking the step change to be identified.\\

\noindent This prompted the development of spatially adaptive smoothing models for spatial data, and the extension to the spatio-temporal domain is one of the contributions of this paper. \citet{brewer2007variable} and \citet{reich2008modeling} extend GMRF models by allowing the variance  $\tau^{2}$ to vary across the study region, while \citet{lawson2012bayesian}, \citet{charras2012classification}, \citet{wakefield2013bayesian} and \citet{anderson2014identifying} utilise a piecewise constant cluster model in the linear predictor to model step changes between neighbouring areas. Alternatively, \citet{lu2007bayesian},  \citet{brezger2007adaptive}, \citet{ma2010hierarchical}, \citet{lee2013locally} and \citet{lee2014bayesian} generalise CAR models by treating the non-zero elements of the adjacency matrix $\mathbf{W}$ as random variables.  Under this approach, equation (\ref{partial}) implies that spatially adjacent $(\phi_{i}, \phi_{k})$ can be partially autocorrelated or conditionally independent, depending on the estimated  value of $w_{ik}$.

\section{Methodology}
\label{sec:model}
\subsection{General approach}
\noindent We present a novel spatially adaptive smoothing model for spatio-temporal data, which allows step changes to occur between adjacent areal units in the unexplained component of risk while treating their locations as unknown.  This is achieved by modelling the adjacency elements in $\mathbf{W}$, that is  $\mathbf{w}^+=\{w_{ik}|i\sim k\}$ (of length $N_{W}=\mathbf{1}^{T}W\mathbf{1}/2$), as random variables on the unit interval, rather than being fixed equal to one. The remaining elements of $\mathbf{W}$ corresponding to non-adjacent areal units remain fixed at zero. Equation (\ref{partial}) shows that when $\rho$ is close to 1, then estimating   $w_{ik}\in\mathbf{w}^+$ as close to 1 results in partial autocorrelation and hence smoothing between the spatially adjacent $(\phi_{ij}, \phi_{kj})$ for all time periods $j$. Conversely, if $w_{ik}$ is estimated as close to zero then $(\phi_{ij}, \phi_{kj})$ are close to conditionally independent for all time periods $j$, and no such spatial smoothing is enforced. In the latter case, a step change is said to exist in the random effects surface between areal units $(i,k)$ for all time periods $j$. We  follow \cite{lu2005bayesian} and quantify the evidence for a step change using

\begin{equation}
p_{ik}=\mathbb{P}(w_{ik}<0.5|\mathbf{Y}),\label{equation pik}
\end{equation}

the posterior probability of $w_{ik}$ being less than 0.5. Our proposed model is one of the first adaptive (localised) smoothing models for spatio-temporal data, and is outlined in two stages below.

\subsection{Level 1 - Likelihood and random effects model for $(Y_{ij}, \phi_{ij})$}
The first level of our proposed model is  given by

\begin{eqnarray}
Y_{ij}|E_{ij}, R_{ij} &\sim & \textrm{Poisson}(E_{ij} R_{ij})~~~~i=1,\ldots,N,~~j=1,\ldots,T,\\
 \ln (R_{ij}) & = & \mathbf{x}_{ij}^{\top}\bd{\beta} +  \phi_{ij},\nonumber\\
 \beta_0 & \sim & \mbox{N}(0, 10000),\nonumber\\
 \bd{\phi}_1 &  \sim & \mbox{N}\left(\bd{0}, \hspace{0.1cm}\tau^2 \mathbf{Q}(\mathbf{W}, \epsilon)^{-1} \right),\nonumber\\
\bd{\phi}_j|\bd{\phi}_{j-1} &  \sim &\mbox{N}\left(\alpha \bd{\phi}_{j-1}, \hspace{0.1cm}\tau^2 \mathbf{Q}(\mathbf{W}, \epsilon)^{-1} \right)  \text{   for   } j=2,\dots,T,\nonumber \\
 \tau^2 & \sim & \mbox{Inverse-Gamma}(0.001,0.001),\nonumber \\
 \alpha & \sim & \mbox{Uniform}(0,1).\nonumber
 \end{eqnarray}

The only difference from the model proposed by  \citet{rushworth2014spatio} is that the GMRF prior proposed by  \citet{Leroux2000} is replaced by the intrinsic GMRF  prior (where $\rho=1$), which is enforced because attempting to estimate $\rho$ and $\mathbf{W}$ could result in high posterior correlation and multimodality, because the random effects are spatially independent if either $\rho=0$ or all elements of $\mathbf{w}^+$ equal zero. To avoid rank-deficiency of the precision matrix and subsequent problems with matrix inversion, the adjusted specification $\mathbf{Q}(\mathbf{W}, \epsilon) = \textrm{diag}(\mathbf{W}\mathbf{1}) - \mathbf{W} + \epsilon \mathbf{I}$ is used, where  $\epsilon \mathbf{I}$ is added to ensure that $\mathbf{Q}(\mathbf{W}, \epsilon)$ is diagonally dominant and hence invertible. This invertibility condition is required because in the second level of the model described below, elements in $\mathbf{W}$ are treated as random variables, necessitating the evaluation of the normalised prior density $f(\bd{\phi}_{j}|\bd{\phi}_{j-1})$.  Sensitivity to the value of $\epsilon$ was checked in an initial modelling step, and was found not to affect estimation until $\epsilon$ was increased to a relatively large value, such as $\epsilon > 10^{-2}$. Therefore in this paper we set $\epsilon=10^{-7}$.

\subsection{Level 2 - Adjacency model for elements in $\mathbf{w}^+$}
\label{sec:level3}
\noindent Our methodological contribution extends the model of \citet{rushworth2014spatio} by treating the elements $\mathbf{w}^+$ as binary random quantities on the unit interval, rather than being fixed at one. Specifying a continuous domain for $\mathbf{w}^{+}$ allows the direct application of a second stage GMRF prior, avoiding the need for a discrete prior such as the Ising model, for which no polynomial time algorithm exists to compute its normalising constant. We model $\mathbf{w}^{+}$ on the logit scale, $\mathbf{v}^+ = \log\left(\mathbf{w}^{+}/(\mathbf{1} - \mathbf{w}^{+}) \right)$, which has the back transformation $\mathbf{w}^{+}= \exp(\mathbf{v}^+) / (1+\exp(\mathbf{v}^+))$. The GMRF prior we propose for $\mathbf{v}^+$ has a constant mean $\mu$, a constant variance $\zeta^{2}$, and a precision matrix defined by the GMRF prior proposed by \citet{Leroux2000}.  This second stage GMRF prior requires us to specify an adjacency structure for the elements in  $\mathbf{v}^{+}$, and here $v_{ik}, v_{rs}\in\mathbf{v}^{+}$ are defined as adjacent (denoted $ik\sim rs$) if the geographical borders they represent in the study region share a common vertex.  Using this notation, we propose the following GMRF prior for $\mathbf{v}^{+}$:

\begin{eqnarray}
p(\mathbf{v}^{+}|\zeta^2, \rho, \mu) & \propto & \exp\left[-\frac{1}{2\zeta^2}\left(\rho\sum_{ik\sim rs}(v_{ik} - v_{rs})^2 + (1-\rho)\sum_{v_{ik}\in\mathbf{v}^{+}}(v_{ik} - \mu)^2\right)\right], \label{eq:edgeprior}\\
\zeta^{2}&\sim&\mbox{Inverse-Gamma}(0.001, 0.001),\nonumber\\
\rho&\sim&\mbox{Uniform}(0, 1).\nonumber
\label{eqn:leroux}
\end{eqnarray}

\noindent This form highlights the role of $\rho$, which controls the extent to which step changes appear spatially clustered together around common vertices. When $\rho\approx 1$ the random variable $v_{ik}$, which controls the existence of a step change between areal units $(i,k)$, is smoothed spatially towards adjacent $v_{rs}$ via the penalty $\sum_{ik\sim rs}(v_{ik} - v_{rs})^2$, which thus induces spatially clustered step changes, this model is conceptually similar to the `CAR2' model proposed in \citet{ma2010hierarchical}. In contrast, when $\rho\approx0$ each $v_{ik}$ is smoothed non-spatially towards the overall mean $\mu$ by the penalty $\sum_{v_{ik}\in\mathbf{v}^{+}}(v_{ik} - \mu)^2$, which does not encourage spatial clustering of step changes. In order to avoid numerical problems when transforming between $\mathbf{v}^{+}$ and $\mathbf{w}^{+}$, the sample space for each $v_{ik}\in\mathbf{v}^{+}$ is truncated to the interval $[-15, 15]$, yielding a sample space of $[0.000000306, 0.9999997]$ for $w_{ik}$, which is close to the intended $[0,1]$ interval.\\

\begin{figure}[h]
\centering
\begin{tabular}{cc}
\includegraphics[width = 6.5cm]{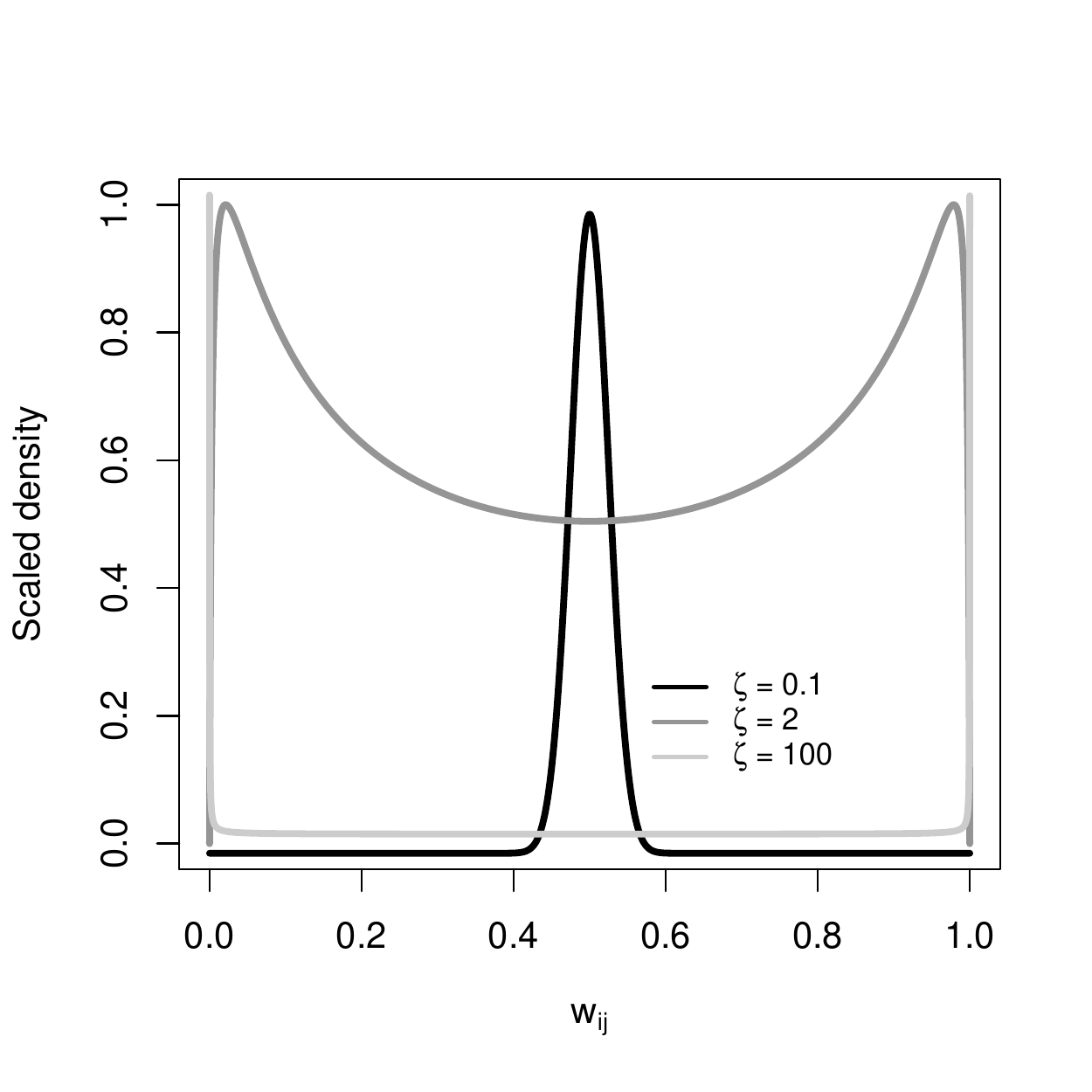}
& \includegraphics[width = 6.5cm]{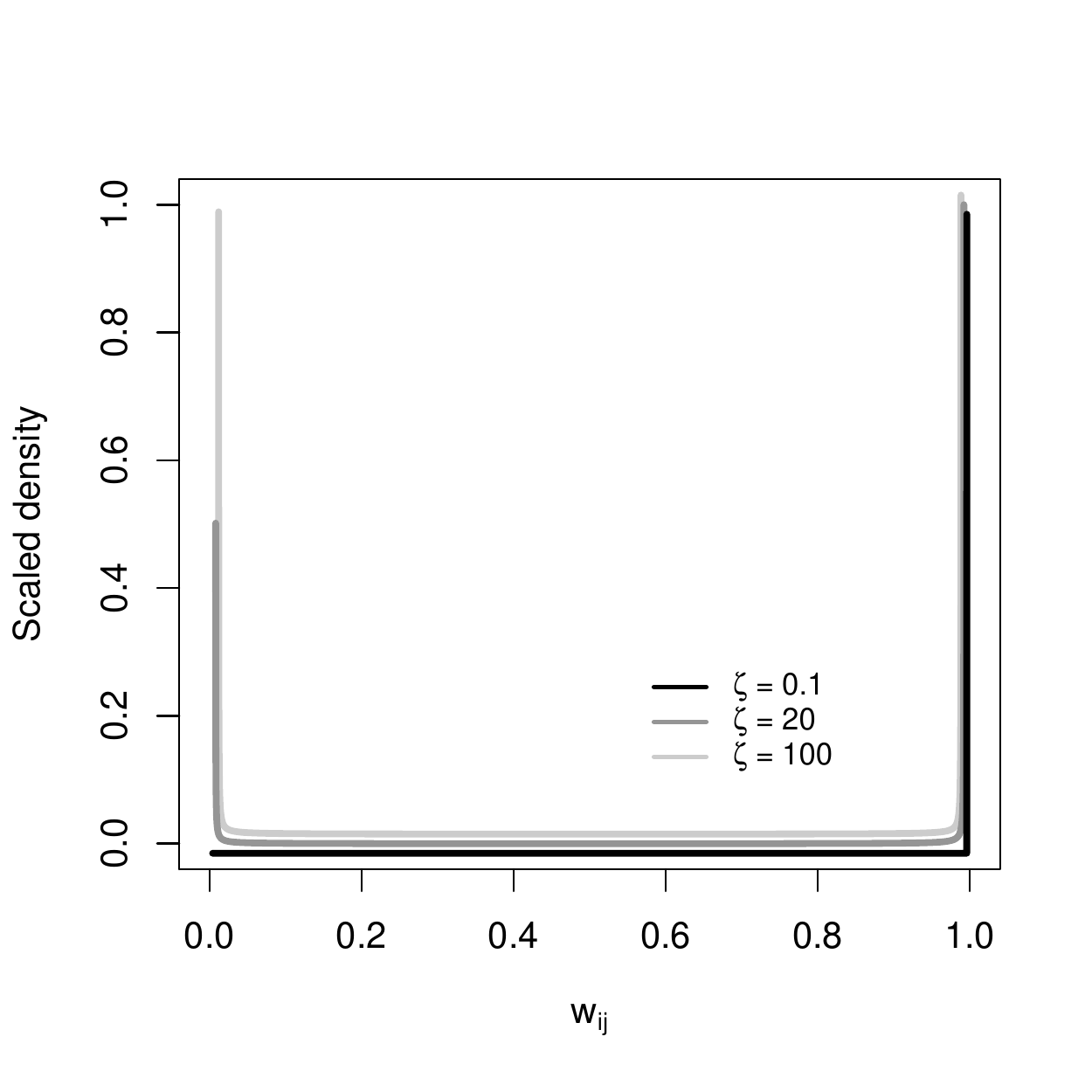}
\end{tabular}
\caption{Plots showing scaled prior densities for $w_{ij}$ for prior means $\mu = 0$ (left) and $\mu = 15$ (right).  In each plot the densities resulting from different $\zeta$ values are shown by different coloured lines.}
\label{fig:bath}
\end{figure}

\noindent The prior mean $\mu$ is fixed in (\ref{eq:edgeprior}), to ensure that the induced prior on the untransformed $\mathbf{w}^{+}$ scale is consistent with our prior beliefs about the prevalence of step changes. Specifically, given the level of spatial autocorrelation evident in the residuals shown in Figure \ref{fig:SIR}, and the associated Moran I statistics reported in Section \ref{sec:motivating}, one would expect there to be relatively few step changes in the random effects surface. In order to be consistent with this preference we choose $\mu>0$, as choosing $\mu < 0$ implies a marginal mean for $w_{ik}$ of less than  $\exp(0)/(1+\exp(0))=0.5$. However, Figure \ref{fig:bath} shows that the induced prior distribution for $w_{ik}$ depends on $\zeta^{2}$ as well as $\mu$, with the left and right panels showing $\mu=0$ and $\mu=15$ respectively for various values of $\zeta$.  When $\mu=0$ the prior density for $w_{ik}$ can have a mode at 0.5, which is incongruous with our prior beliefs about $w_{ik}$ being close to one for most $w_{ik}\in \mathbf{w}^{+}$. Initial simulations confirmed that setting $\mu=0$ leads to spurious step changes being identified. In contrast, when $\mu=15$ the prior assigns high prior probability to $w_{ik}\approx 0$ or $w_{ik}\approx 1$ or both, with little prior probability in between. The ratio of the densities at $\{0,1\}$ depends on $\zeta$, so that when $\zeta$ is small, almost all prior mass is concentrated around $w_{ik} = 1$, hence strongly discouraging boundaries. In contrast, as $\zeta$ increases the prior becomes more symmetric and  `U' shaped, with equal point masses at 0 and 1 expressing ambivalence about the presence or absence of step changes. Thus fixing $\mu=15$ ensures that clear step change decisions, that is $w_{ik}$ close to zero or one, are preferred over ambiguous values such as $w_{ik}=0.5$.

\subsection{Inference}

\noindent Our model is fitted within a Bayesian framework using Markov-chain Monte Carlo (McMC) sampling, and is freely available via the \texttt{R} package \texttt{CARBayesST} (\citet{lee2015carbayesst}). The McMC algorithm is a combination of Gibbs and Metropolis-Hastings steps, and the high dimensionality of the parameter space requires an algorithm that minimises the overall computational burden.  We exploit matrix sparsity resulting from the GMRF prior precision matrix wherever possible, using efficient loops written in C++ and matrix triplet form to perform the algebraic manipulations.  The random effects were updated using fast one-at-a-time Metropolis-Hastings steps rather than using a joint or block updating scheme.  While joint updates (for example  \citet{knorr2002block}) can achieve substantially better mixing for an equivalent number of McMC steps and particularly when the parameters are highly correlated, the computational cost was found to outweigh the benefits of better sampling efficiency. Further details are given in the supporting information accompanying this paper.

%%%%%%%%%%%%%%%%%%%%%%%%%%%%%%%%%%%%%%%%%%%%%%%%%%
%%%%%%%%%% SIMULATION STUDY
%%%%%%%%%%%%%%%%%%%%%%%%%%%%%%%%%%%%%%%%%%%%%%%%%%
\section{Simulation study}
\label{sec:simulation}

\noindent In this section we comprehensively test the performance of two variants of  the proposed model on simulated data under a range of scenarios,  and compare their performance against two commonly used alternatives. The two existing models are those proposed by \citet{knorr2000bayesianmodelling} (denoted \textbf{(1)})  and \citet{rushworth2014spatio} (denoted \textbf{(2)}), although the former is implemented with GMRF priors proposed by \citet{Leroux2000} rather than a convolution of independent and intrinsic GMRF priors. Additionally, model \textbf{(1)} is implemented with independent interactions (Gaussian with zero-mean and a common variance), although 3 others types of interactions were proposed by \citet{knorr2000bayesianmodelling}. For more details see the vignette accompanying the \texttt{CARBayesST} software. Model \textbf{(3)} is the adaptive smoothing model proposed here with the simplification that  $\rho = 0$, while  Model \textbf{(4)} is the full model where $\rho$ is not treated as fixed. Model \textbf{(3)} \emph{a-priori} treats each $w_{ik}\in\mathbf{w}^{+}$ independently, and therefore does not encourage configurations in which step changes cluster together. Our primary focus in this study is to assess the ability of each model to estimate the spatio-temporal pattern in disease risk, and identify step changes in risk between neighbouring areas.

\subsection{Data generation and study design}
\begin{figure}[h]
\centering
\begin{tabular}{cc}
 \includegraphics[clip, trim = 10mm 72mm 10mm 72mm, width = 8cm, angle = 270]{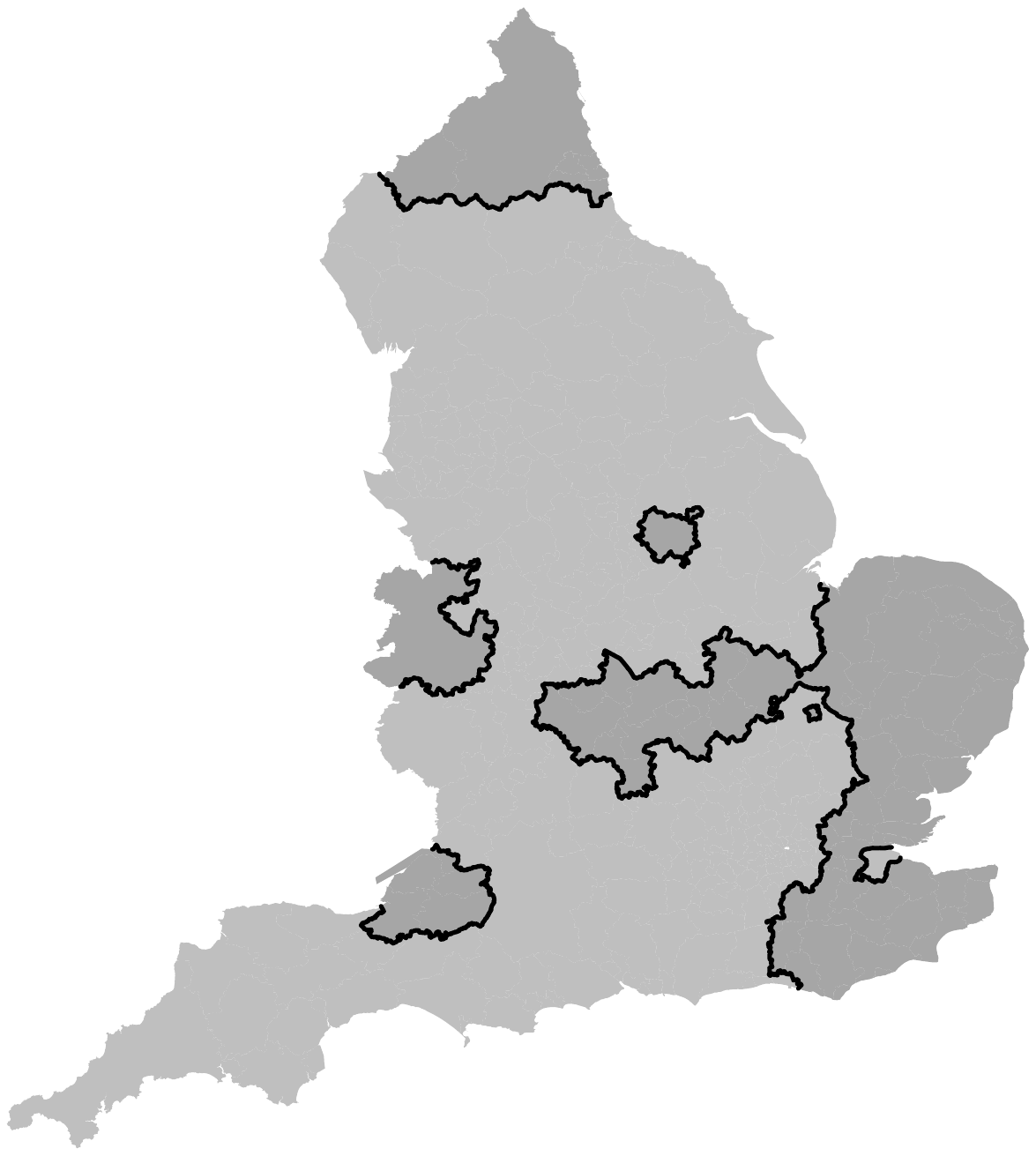}
& \hspace{0cm}\includegraphics[clip, trim = 10mm 72mm 10mm 72mm, width = 8cm, angle = 270]{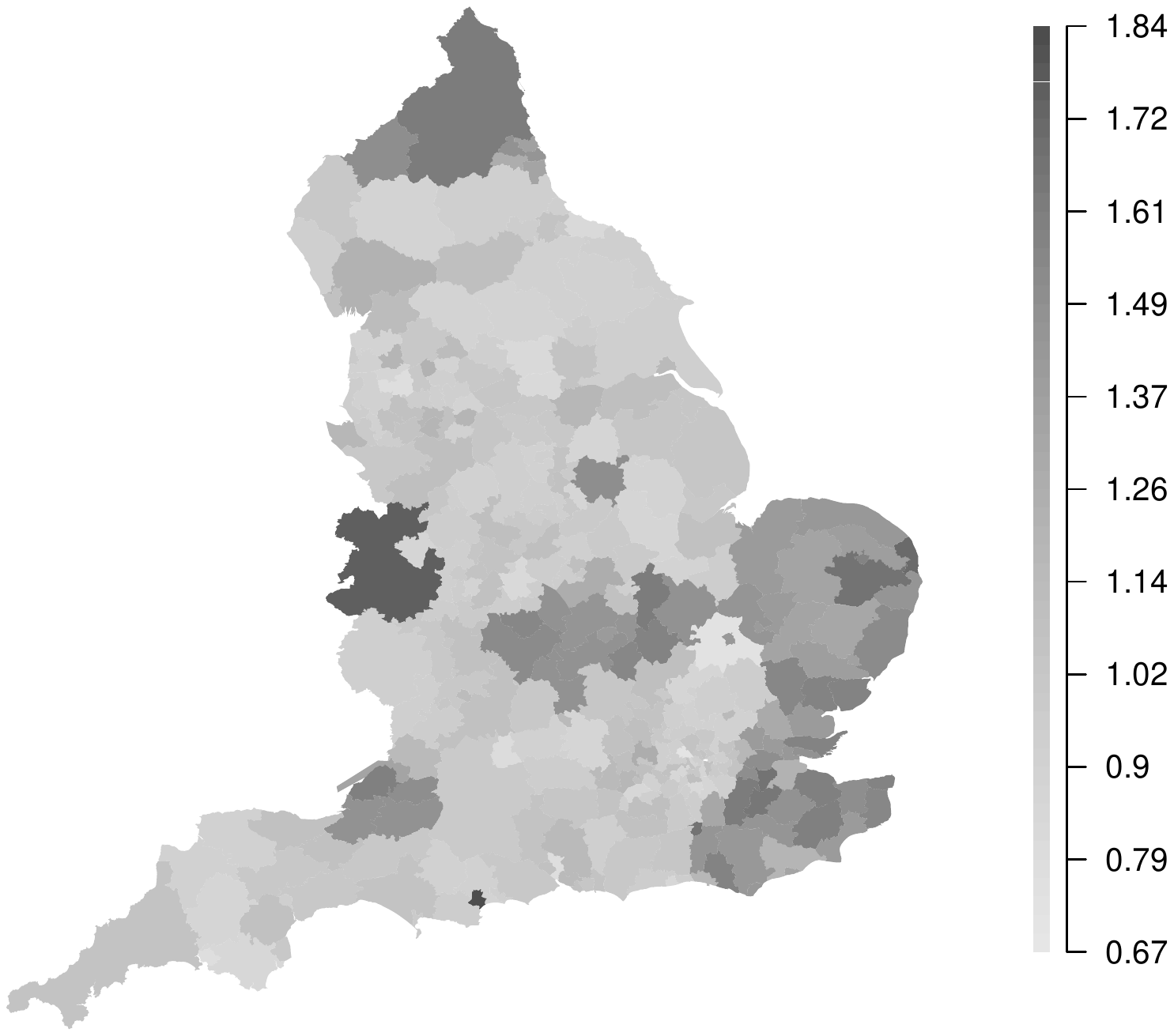} \\
\end{tabular}
\caption{Left: Locations of the true step changes in risk, illustrated by black lines following the borders between the selected subregions.  Darker shading indicates areas with true risk of 1.5 while lighter shading indicates a true risk of 1.  Right: A single realisation of the spatial risk surface assuming $A=1.5$.}
\label{fig:spatialstructure}
\end{figure}

\noindent  Simulated disease counts $\{Y_{ij}\}$ are generated for the England study region from a Poisson log-linear model, that is $Y_{ij}\sim\mbox{Poisson}(E_{ij}R_{ij})$, where both the size of the expected counts and the number of time periods $T$ are varied to assess their impact on model performance. The log-risk surfaces are generated for each time period from a multivariate Gaussian distribution, whose precision matrix is defined by the intrinsic GMRF prior (\citealp{besag1974spatial}, \citealp{besag1991bayesian}) and hence produces spatially smooth surfaces. To simulate spatial step changes in log-risk, a piecewise constant mean surface is specified for the random effects,  which is displayed in the left panel of Figure \ref{fig:spatialstructure}. Lighter shaded areas exhibit a mean risk of 1 while the darker shaded areas have a mean risk level of $A$, and the black lines correspond to the locations of true step changes. An example realisation of this surface is shown in the right panel of Figure \ref{fig:spatialstructure} for $A=1.5$, where the clusters of high-risk areas are evident. To ensure the true risk surface is not identical for all time periods, independent random noise is added to the risk in each areal unit for each time period. The scenarios considered in this study are summarised in Table \ref{table:simulationchoices}, which shows that we consider $T=1, 5, 20$ time periods,  elevated risk levels of $A=1,1.5,2$, and disease prevalences of $E=10,50,100$. For the $A=1$ scenario this corresponds to a spatially smooth risk surface with no step changes, which tests the model's propensity for identifying step changes when none are present (false positives).

  \begin{table}
   % \label{table:model}
    \caption{\label{table:simulationchoices}Description of the scenarios in the simulation study.}
    \vspace{0.2cm}
  \centering
\begin{tabular}{l  l  l}
\hline  \textbf{Scenario type} & \textbf{Parameters varied} & \textbf{Parameters fixed} \\ 
    \hline
    Varying time dimension & $T \in \{1, 5, 20 \}$ & $a = 1.5$; $E =  75$ \\ 
    Relative risk in high regions &  $A \in \{1, 1.5, 2 \}$ & $T = 5$; $E = 75$ \\ 
    Expected cases & $E \in \{10, 50, 100 \}$  & $T = 5$; $a = 1.5$ \\\hline
\end{tabular}
\end{table}

\subsection{Results}
\label{sec:simresult}
\noindent  One hundred data sets were generated under each of the 9 scenarios shown in Table \ref{table:simulationchoices}, and Models \textbf{(1)} - \textbf{(4)} were fitted in turn. Inference for each model was based on 30,000 McMC samples following a burn-in period of 20,000 samples, after which convergence was assessed to have been reached.  The quality of the estimation of the spatio-temporal pattern in disease risk was quantified by its root-mean squared error, RMSE$=\sqrt{\frac{1}{NT}\sum_{i,j}(R_{ij}-\hat{R}_{ij})^{2}}$, as well as by the Deviance Information Criterion (DIC), effective number of parameters ($pD$) and the coverage probabilities of the 95$\%$ credible intervals. Receiver-operating characteristic (ROC) curves were also computed to quantify the accuracy of the step change detection, which were based on each models sensitivity and specificity at identifying true step changes. These statistics compared $\mathbb{E}[w_{ij}| \mathbf{Y}]$ to a threshold value $p^{*}$, where if $\mathbb{E}[w_{ij}| \mathbf{Y}]<p^{*}$ a step change was identified where as for the converse no step change was declared. The value of $p^{*}$ was varied from 0 to 1 at intervals of 0.01, and the ROC curve is a plot of sensitivity against specificity. However, for ease of presentation the Area Under the Curve (AUC) is presented here rather than the full ROC curve, and an AUC of one corresponds to perfect step change identification. AUC associated with step-change estimation was only available for Models \textbf{(3)} and \textbf{(4)}, as Models \textbf{(1)} and \textbf{(2)} do not estimate step-changes as $\mathbf{w}^+$ is fixed and not estimated in these models.  Therefore, AUC is of interest to validate and compare the performance the adaptive models proposed.\\

\begin{center}
\begin{table}
\caption{\label{table:rmsetable1} Median root mean-squared error (RMSE), 95\% credible interval coverages associated with the fitted risks, Deviance Information Criterion (DIC), and the effective number of parameters (pD)  for each model and scenario.}
\begin{tabular}{llccccccccc}
\hline \\
 && \multicolumn{4}{ c }{\textbf{RMSE}} && \multicolumn{4}{ c }{\textbf{DIC}}\\
 && \textbf{(1)} & \textbf{(2)} & \textbf{(3)} & \textbf{(4)} && \textbf{(1)} & \textbf{(2)} & \textbf{(3)} & \textbf{(4)}\\
  \hline
 \multirow{3}{*}{\textbf{Time}} 
  & $T=1$  & 0.092 & 0.092 & 0.088 & 0.090 &&  2497 & 2497 & 2489 & 2502 \\ 
  & $T=5$  & 0.052 & 0.070 & 0.034 & 0.062 && 11986 & 12116 & 11815 & 12132\\ 
  & $T=20$ & 0.036 & 0.058 & 0.032 & 0.036 && 47355 & 47690 & 47217 & 47325\\ 
  \hline
  \multirow{3}{*}{\textbf{Risk}} 
  & $A=1$   & 0.023 & 0.027 & 0.024 & 0.024 && 11598 & 11625 & 11608 & 11614 \\ 
  & $A=1.5$ & 0.052 & 0.071 & 0.035 & 0.062 && 11983 & 12116 & 11815 & 12107 \\ 
  & $A=2$   & 0.059 & 0.089 & 0.039 & 0.059 && 12131 & 12442 & 11923 & 12200 \\ 
  \hline
  \multirow{3}{*}{\textbf{Cases}} 
  & $E=10$  & 0.100 & 0.129 & 0.111 & 0.114 && 8549 & 8570 & 8577 & 8595 \\ 
  & $E=50$  & 0.061 & 0.079 & 0.043 & 0.069 && 11293 & 11393 & 11164 & 11407 \\ 
  & $E=100$ & 0.046 & 0.063 & 0.031 & 0.051 && 12472 & 12634 & 12284 & 12575\\ \hline
  \hline \\
 && \multicolumn{4}{ c }{\textbf{pD}} && \multicolumn{4}{ c }{\textbf{Coverage}}\\
 && \textbf{(1)} & \textbf{(2)} & \textbf{(3)} & \textbf{(4)} && \textbf{(1)} & \textbf{(2)} & \textbf{(3)} & \textbf{(4)}\\
  \hline
 \multirow{3}{*}{\textbf{Time}} 
  & $T=1$  & 166 & 166 & 155 & 164  &&   0.930 & 0.930 & 0.955 & 0.950 \\ 
  & $T=5$  & 322 & 526 & 159 & 520  &&   0.970 & 0.940 & 0.980 & 0.960\\ 
  & $T=20$ & 565 & 1322 & 351 & 377 &&   0.960 & 0.910 & 0.910 & 0.850\\ 
  \hline
  \multirow{3}{*}{\textbf{Risk}} 
  & $A=1$   & 98 & 234 & 36 & 5 && 0.986 & 0.996 & 0.878 & 0.263 \\ 
  & $A=1.5$ & 323 & 528 & 164 & 475 && 0.963 & 0.930 & 0.979 & 0.956 \\ 
  & $A=2$   & 363 & 761 & 173 & 475 && 0.963 & 0.941 & 0.982 & 0.964 \\ 
  \hline
  \multirow{3}{*}{\textbf{Cases}} 
  & $E=10$  & 159 & 274 & 186 & 201 && 0.964 & 0.941 & 0.950 & 0.944 \\ 
  & $E=50$  & 290 & 456 & 154 & 413 && 0.962 & 0.932 & 0.985 & 0.963 \\ 
  & $E=100$ & 347 & 584 & 171 & 480 && 0.965 & 0.936 & 0.978 & 0.956\\ \hline
\end{tabular}
\end{table}
\end{center}

\noindent Table \ref{table:rmsetable1} shows the RMSE, DIC, $pD$  and coverages associated with each model across the nine simulation scenarios, from which a number of patterns emerge.  Firstly, the simplified adaptive model \textbf{(3)} with no spatial smoothing across step changes generally outperforms the non-adaptive models \textbf{(1)} and \textbf{(2)} and the full adaptive model \textbf{(4)} in terms of lower RMSE, DIC and $pD$.  In the $E=10$ scenario model \textbf{(1)} clearly outperforms \textbf{(3)} and  \textbf{(4)} in terms of RMSE, DIC and coverage which is likely the result of incorrect boundary identification from the adaptive models (see \ref{table:roctable1}). Also in the $A=1$ scenario with no true step changes model \textbf{(4)} exhibits a very low $pD$, which indicates oversmoothing caused by false identification of step changes in the surface (see \ref{table:roctable1}).  Overall, Table \ref{table:rmsetable1} strongly suggests that the spatial smoothing imposed on $\mathbf{w}^+$ by \textbf{(4)} is sub-optimal compared to assuming each element $w_{ik}\in\mathbf{w}^+$ is \emph{a-priori} independent as in \textbf{(3)}.  For all models, RMSE decreases as both the number of time periods $T$ and disease prevalence $E$ increases, which is due to an increase in the amount of data. Confidence interval coverage was generally very good, with all coverage levels varying between 93\% and 97\%,  excepting $A = 1$ when the coverages were 0.986, 0.996, 0.878 and 0.263 for models \textbf{(1)}, \textbf{(2)}, \textbf{(3)} and \textbf{(4)}, respectively.\\

 \begin{table}
    \caption{\label{table:roctable1}Median area under the ROC curve for step change identification across 100 simulations for models \textbf{(3)} and \textbf{(4)}.  Bracketed figures correspond to the 10\% quantile of areas.  For $A=1$ $\textrm{SPF}$ denotes the specificity since there are no true step changes to identify in this scenario.}
    \vspace{0.2cm}
  \centering
\begin{tabular}{l l c  c }
\hline &&  \multicolumn{2}{c}{\textbf{Median area under ROC curves}} \\
  && \textbf{(3)} & \textbf{(4)}  \\ 
  \hline
 \multirow{3}{*}{\textbf{Temporal} \textbf{replication}} & $T=1$  & 0.7399 (0.6696) & 0.4995 (0.4898)  \\ 
  & $T=5$  & 0.9999 (0.9996) & 0.4995 (0.4988)\\ 
  & $T=20$ &  0.9997 (0.9988) & 0.9151 (0.4995)\\
  \hline
    \multirow{3}{*}{\textbf{Relative risk}} &  $A=1, \hspace{0.1cm}\textrm{SPF}$  & 0.9769 (0.9463) & 0.4186 (0.2649)\\ 
  & $A=1.5$ & 0.9993 (0.9996) & 0.5913 (0.4988)\\ 
  & $A=2$  & 0.9999 (0.9997) & 0.7746 (0.4995)\\  
   \hline
  \multirow{3}{*}{\textbf{Expected cases}} & $E=10$  & 0.6672 (0.6262) & 0.4979 (0.4885)\\ 
  & $E=50$  & 0.9925 (0.9780) & 0.5801 (0.4988)\\ 
  & $E=100$  & 0.9999 (0.9998) & 0.6463 (0.4994)\\ \hline
\end{tabular}
\end{table}

\noindent Table \ref{table:roctable1} displays the median AUC statistic across the set of ROC curves calculated for each scenario. The numbers in brackets are the tenth percentile of that distribution, and summarise the variation across  the 100 simulated data sets. An exception to this is the $A=1$ scenario, which displays the specificity because as the risk surface is spatially smooth there are no true step changes to identify. For model \textbf{(3)} the median AUC values are close to the maximal value of 1, indicating very accurate step change identification. The exception to this occurs when $T = 1$ (AUC = 0.7399) and $E = 10$ (AUC = 0.6672), which results from limited information about step change location provided by the data in both cases. In contrast, Model \textbf{(4)} always performs much less well, with median AUC values ranging between 0.5 and 0.9151. The poor performance of Model \textbf{(4)} is also evident in the tenth percentile results, and re-enforces the RMSE and coverage results displayed in Table \ref{table:rmsetable1}. It is likely to be because model \textbf{(4)} forces the step changes to be spatially smooth, thus inducing a set of false positives that are spatially close to the real boundaries. The other main result from Table \ref{table:roctable1} is that the AUC increases as the number of time periods $T$ increases and as the size of the expected cases increases, which is due to an increase in the amount of data.

%%%%%%%%%%%%%%%%%%%%%%%%%%%%%%%%%%%%%%%%%%%%%%%%%%
%%%%%%%%%% England DATA EXAMPLE
%%%%%%%%%%%%%%%%%%%%%%%%%%%%%%%%%%%%%%%%%%%%%%%%%%

\section{Results of the England case study}
\label{sec:application}

% data description
% data description
\noindent The simulation study has shown that the simplified adaptive model with $\rho=0$ (Model \textbf{(3)}) is capable of identifying spatial discontinuities where they exist, where as the full model (Model \textbf{(4)}) is not, Additionally Model \textbf{(3)} fits the data better than Model \textbf{(4)} in terms of DIC, and thus the latter is not considered here. We thus apply three models to the England circulatory and respiratory data sets, the interaction model of \citet{knorr2000bayesianmodelling} (Model \textbf{(1)}), the global smoothing model of \citet{rushworth2014spatio} (Model \textbf{(2)}) and the adaptive smoothing model proposed here with the simplification that $\rho=0$ (Model \textbf{(3)}). The three covariates discussed in Section 2 are included in each model, which are the proportion of working age people claiming Job Seekers Allowance (JSA), the proportion of each LUA classified as urban (Urbanity) and the particulate matter concentrations (PM$_{10}$). Inference for all models is based on thinning (by 10) $10^6$ posterior samples including a burn-in period of a further $10^5$ samples. In analysing these data our goals are to: (i) estimate the spatio-temporal pattern in disease risk to quantify the extent of health inequalities; and (ii) estimate the location of any step changes in the unexplained spatial risk structure, which will assist in the identification of unmeasured confounders. The analysis provided here can be reproduced using the code and data in the supporting material accompanying this paper.

\subsection{Model fit}
The top two rows of Table \ref{tab:EnglandTable} display the overall fit of each model to each data set, by presenting the DIC  and the effective number of parameters $pD$. It shows that the adaptive smoothing model \textbf{(3)} fits the data better than the global smoothing models \textbf{(1)}, \textbf{(2)} for both diseases, with reductions in DIC in both cases.  Additionally, model \textbf{(3)} has a markedly smaller number of effective parameters ($pD$) than models \textbf{(1)}, \textbf{(2)}, despite having a more complex specification. This is because its ability to identify step changes permits the GMRF component to smooth more strongly elsewhere in the spatial surface, resulting in smaller variance estimates for $\tau^{2}$ than from model \textbf{(2)}. This implies a greater level of penalisation of the random effects and hence a reduction in the overall $pD$. 

  \begin{table} 
    \caption{\label{tab:EnglandTable} Diagnostics for models  \textbf{(1)} to \textbf{(3)} for the England circulatory and respiratory  admissions data sets.}
  \centering
\begin{tabular}{ccccccc}
\hline&  \multicolumn{3}{c}{\textbf{Circulatory}} &  \multicolumn{3}{c}{\textbf{Respiratory}}  \\
\textbf{Diagnostic} & \textbf{(1)} &  \textbf{(2)} & \textbf{(3)} &  \textbf{(1)} & \textbf{(2)} & \textbf{(3)}  \\
  \hline
 DIC & 35,269 & 35,314 & 35,056 &  35,223 & 35,239 & 34,993\\ 
    $pD$ & 2,783 & 2,890 & 2630 & 2,825 & 2,887 & 2,629 \\
 $\%$ of borders with $p_{ik}>0.75$ & - & - &  30.8 & - & - & 32.6\\
  $\%$ of borders with $p_{ik}>0.99$ & - & - & 14.9  & - & - & 17.1\\  
 $\hat{\tau}^2$ & - & 0.0295 & 0.0127 &  - & 0.0395 & 0.0130\\
  $\hat{\zeta}^2$ & - & - & 250.4 & - & - & 254.9\\
   $\hat{\alpha}$ & - & 0.963 & 0.961 &  - & 0.969 & 0.964\\\hline
\end{tabular}
\end{table}

\subsection{Covariate effects}
The effects of the three covariates on circulatory and respiratory disease risks estimated by model \textbf{(3)}, the best fitting model, are displayed in Table \ref{tab:coeffTable}, where all results are presented as relative risks associated with one standard deviation increases in each covariate.  Standard deviations were calculated using the raw covariate data, and are shown in the second column of Table \ref{tab:coeffTable}.  The table shows that increasing socio-economic deprivation (as measured by JSA) increases risk for hospital admission of both diseases, with relative risks of 1.08 (circulatory) and 1.135 (respiratory) respectively. This harmful effect of poverty is well known, and is a proxy for differences in average lifestyle such as prevalence of smoking, drinking, exercise etc. Both diseases also show substantial effects of urbanicity on disease risk, with more urban areas exhibiting increased risks of 1.096 (circulatory) and 1.142 (respiratory) respectively. Finally, particulate matter air pollution appears to have no impact on circulatory disease risk and a slight effect on respiratory disease risk, with the latter relative risk being 1.021.

\begin{table} 
\caption{\label{tab:coeffTable} Relative risks and associated 95\% credible intervals associated with 1 standard deviation increases in each covariate, as estimated by model \textbf{(3)}}
\centering
\begin{tabular}{cccc}
\hline & \multicolumn{3}{c}{\textbf{Circulatory}} \\
\textbf{Covariate} & \textbf{St.Dev} & \textbf{RR} & \textbf{95\% CI} \\ 
\hline
\textbf{JSA}   & 0.0476 & 1.0799 & (1.0586, 1.1021) \\ 
\textbf{Urbanicity}  & 0.257 & 1.0964 & (1.0757, 1.1173)\\
\textbf{PM}$_{10}$  & 2.903 & 0.9954 & (0.9730, 1.0209)\\ 
\hline & \multicolumn{3}{c}{\textbf{Respiratory}} \\
\textbf{Covariate} & \textbf{St.Dev} & \textbf{RR} & \textbf{95\% CI} \\ 
\hline
\textbf{JSA}   & 0.0476 & 1.1346  & (1.1126, 1.1555) \\ 
\textbf{Urbanicity}  & 0.257 & 1.1418 & (1.1217, 1.1649)\\
\textbf{PM}$_{10}$  & 2.903 & 1.0211 & (0.9946, 1.0557)\\\hline
\end{tabular}
\end{table}

\subsection{Health inequalities}
Maps of the average risks across all years from model \textbf{(3)} are displayed in the left column of Figure \ref{fig:EnglandBoundaries}, and show similar spatial patterns in risk for both diseases, with a Pearson's correlation coefficient of 0.892. The maps show that the average risk varies over space with values between (0.433, 1.636) and (0.175, 2.147) respectively for circulatory and respiratory disease, suggesting the presence of substantial health inequalities. These inequalities have generally widened over time, as the difference between the highest and lowest respiratory disease risk was 1.77 in 2001 and 2.13 in 2010. For circulatory disease a similar pattern is evident, with an estimated difference between highest and lowest risk of 1.39 in 2001 and 1.54 in 2010.

\subsection{Step change identification}
\begin{figure}[h]
\begin{tabular}{cc}
 \includegraphics[clip, trim = 50mm 30mm 60mm 30mm, width = 6.4cm]{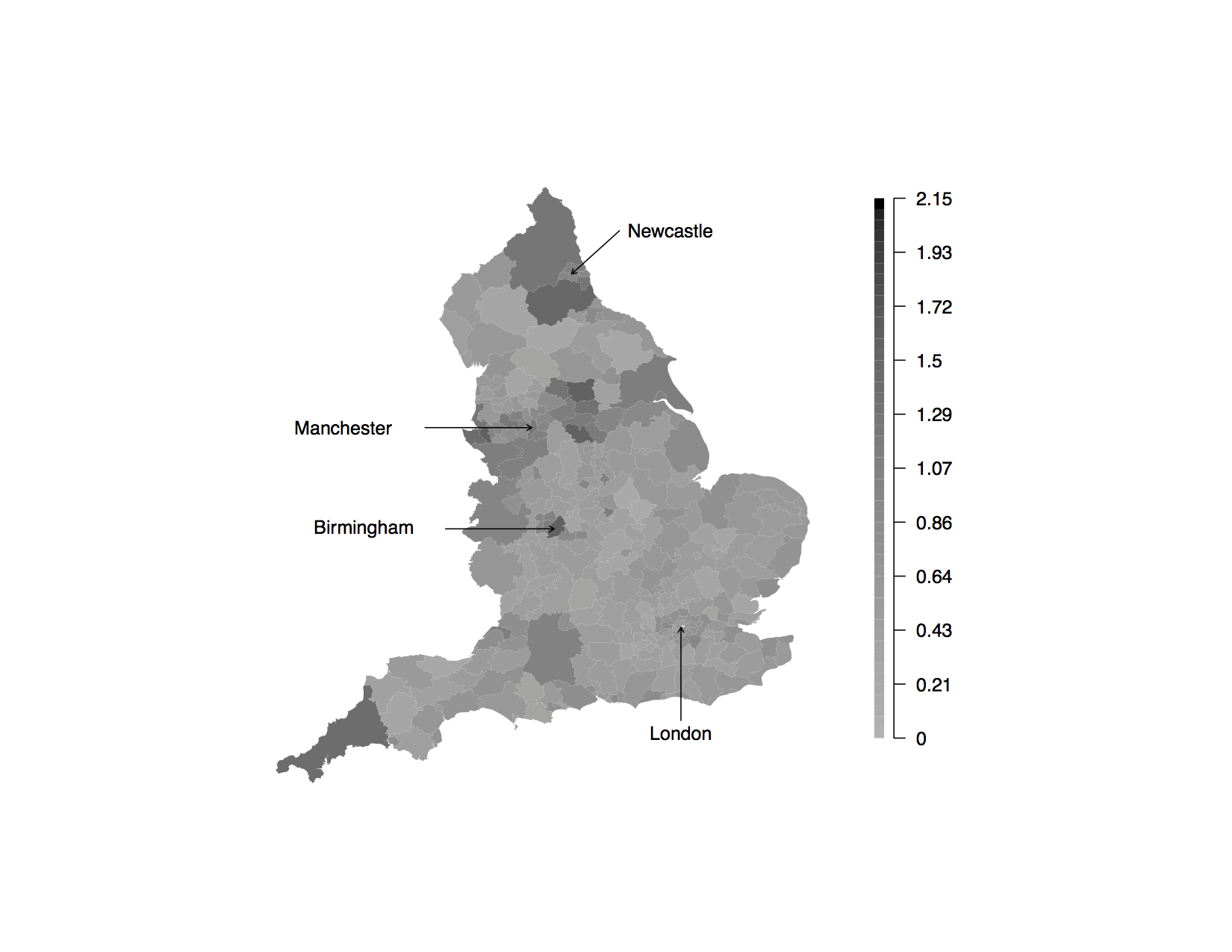}
& \includegraphics[clip, trim = 50mm 30mm 60mm 30mm, width = 6.4cm]{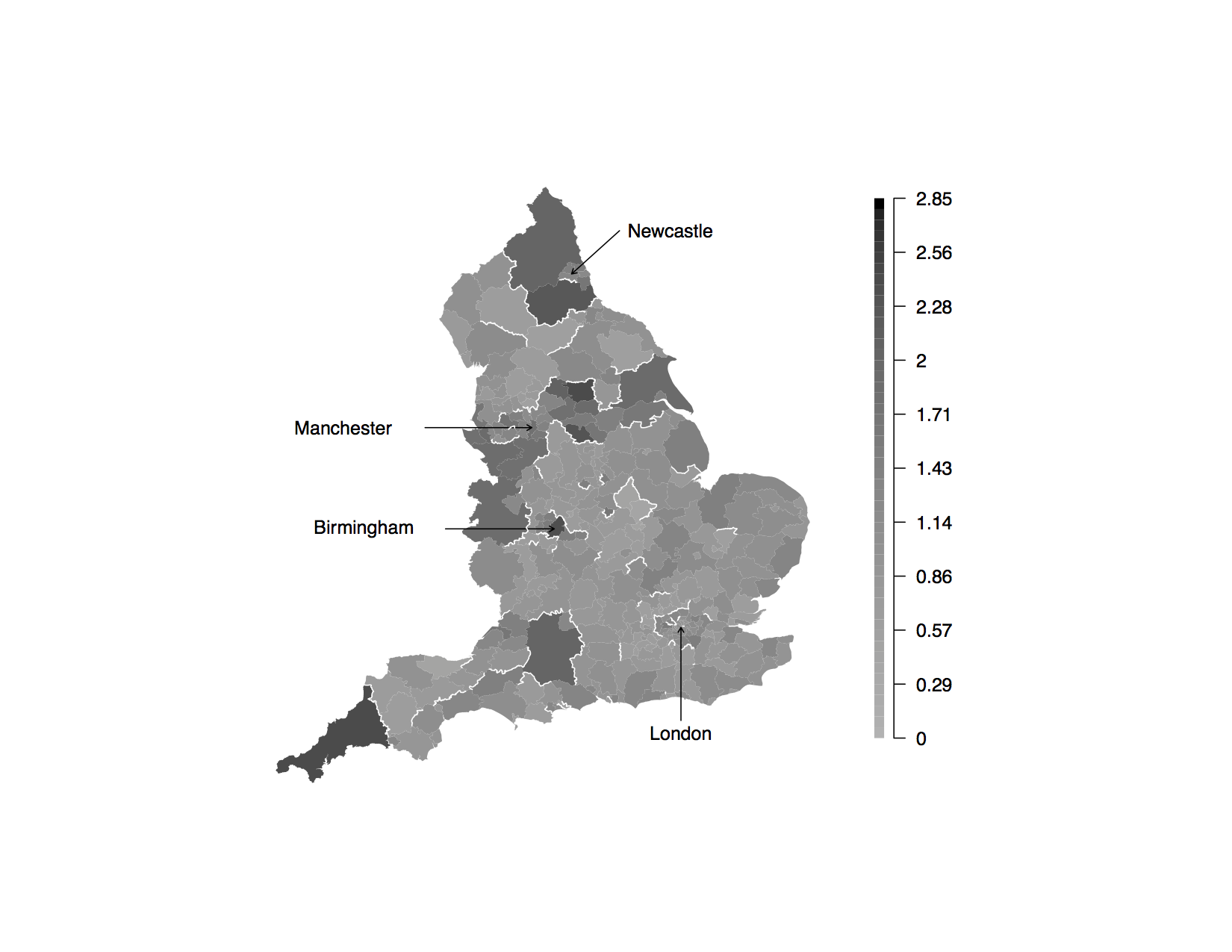} \\
 \includegraphics[clip, trim = 50mm 30mm 60mm 30mm, width = 6.4cm]{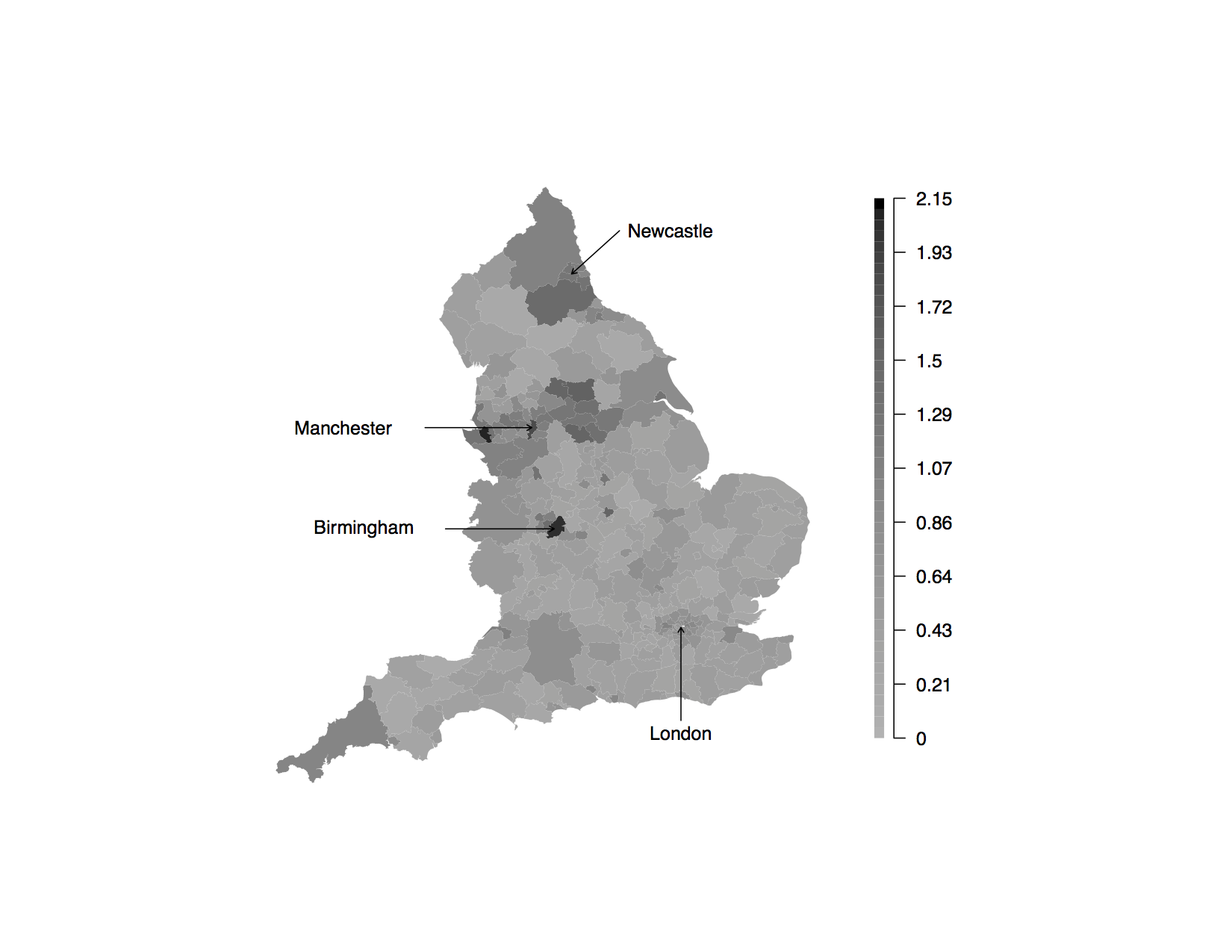}
& \includegraphics[clip, trim = 50mm 30mm 60mm 30mm, width = 6.4cm]{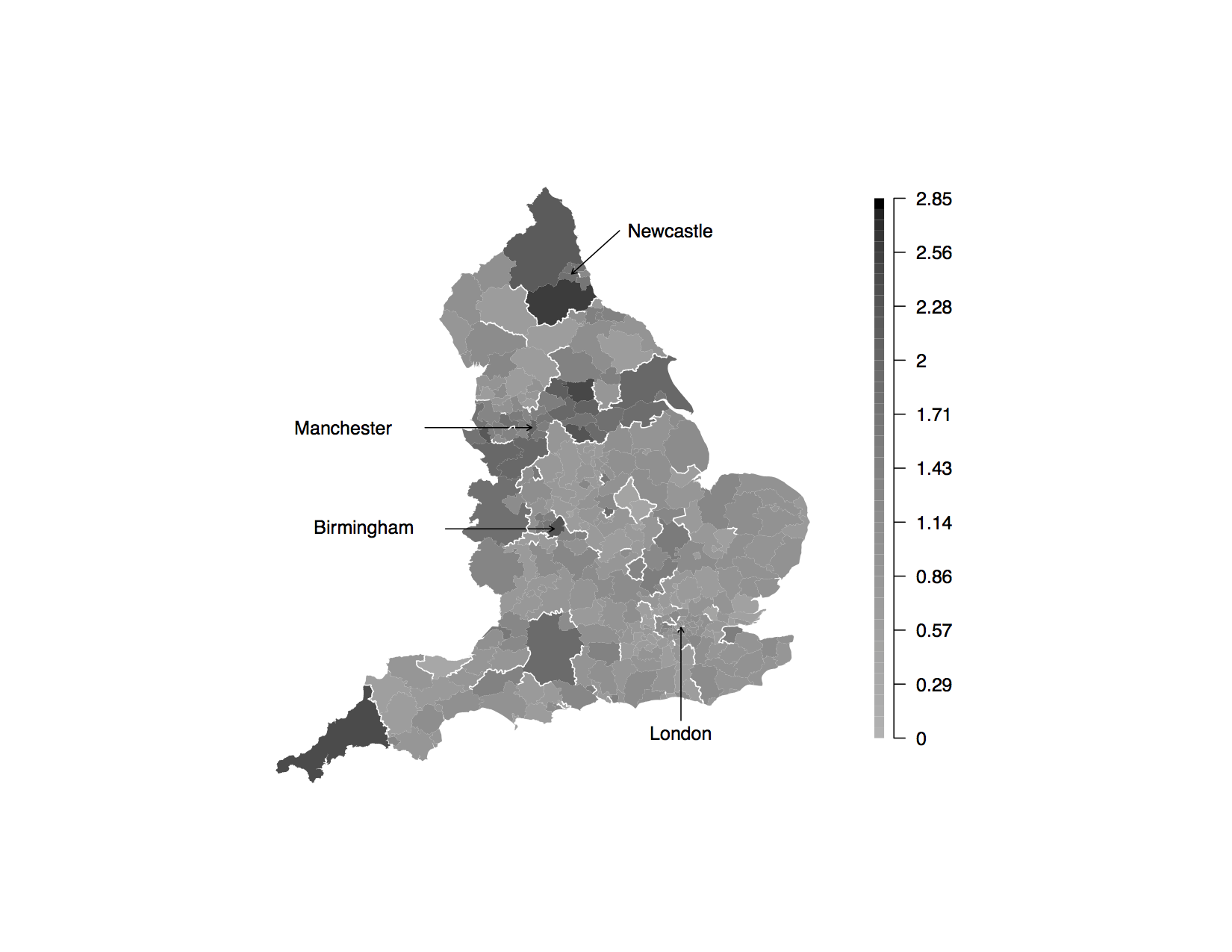} \\
\end{tabular}
\caption{Maps showing the average risk surface (left column) and the unexplained component of the risk surface (right column) for both diseases from model \textbf{(3)}. The top row relates to circulatory disease while the bottom row relates to respiratory disease. The white lines on the maps in the right column correspond to step changes that have been identified using a cutoff of $p_{ik}\geq 0.99$ in (\ref{equation pik}).}
\label{fig:EnglandBoundaries}
\end{figure}

\noindent Table \ref{tab:EnglandTable} summarises the number of step changes in the unexplained component of the risk surface, based on $p_{ik}=\mathbb{P}(w_{ik}<0.5|\mathbf{Y})$ values above a threshold of 0.75 and 0.99. The higher 0.99 level threshold was used by \cite{lu2005bayesian}, and results in 15.1$\%$ of borders being step changes for circulatory disease and 17.1$\%$ for respiratory disease. These step changes are largely similar between the diseases, with 92$\%$ agreement between their locations. They are displayed in the right column of Figure \ref{fig:EnglandBoundaries} as white lines, while the grey shading represents the time averaged exponentiated random effects surface  which corresponds to the unexplained component of the variation in disease risk. The figure shows evidence of much higher unexplained risks of hospital admission in areal units containing large cities, and in the central band of Northern England that incorporates Manchester and Yorkshire, even after adjusting for the covariates. It is striking that these features are largely consistent between the two diseases, so that although the estimated risks have different overall magnitudes, they exhibit very similar spatial patterns. Public health professionals can use these results to identify potential risk factors for disease, by searching for risk factors that exhibit step changes in the same locations as those exhibited in Figure \ref{fig:EnglandBoundaries}.

\subsection{Sensitivity analysis}
A prior sensitivity analysis was performed to test the robustness of the above results to changes in prior specifications.  In particular, the shape and scale parameters in the hyperpriors for $\zeta^2$ and $\tau^2$ were changed to (1,1) and (1, 0.001), and the prior variance for $\bd{\beta}$ was reduced to 100.  The results reported above were found to be robust to these modifications.  Our choice of Inverse-Gamma priors here is due to their conjugacy, but alternatives could be explored such as the  half-Cauchy distribution as suggested by \citet{gelman2006prior}.

%%%%%%%%%%%%%%%%%%%%%%%%%%%%%%%%%%%%%%%%%%%%%%%%%%
%%%%%%%%%% CONCLUSIONS
%%%%%%%%%%%%%%%%%%%%%%%%%%%%%%%%%%%%%%%%%%%%%%%%%%

\section{Discussion}
\label{sec:discussion}
\noindent In this paper a new study of the spatio-temporal structure of circulatory and respiratory disease risk in England was presented, with the goal of understanding the extent of health inequalities and whether the data present evidence of step changes in disease risk between pairs of adjacent regions.  Consequently, a new spatially adaptive smoothing model was developed, that can estimate the location and strength of such step-changes. The model is a spatially adaptive extension to the class of GMRF prior distributions, and is one of the first models for step change identification in spatio-temporal disease risk. Freely available software via the \texttt{CARBayesST} package for \texttt{R} is provided to allow others to utilise our  model, and this is one of the first \texttt{R} packages for spatio-temporal areal unit modelling.\\

\noindent The simulation study in Section 5 established the superiority of our model over global smoothing alternatives, in terms of both risk estimation and model fit.  Our model was successful  at recovering the locations of known step changes in simulated data, with AUC statistics close to one for a range of different scenarios. These AUC statistics were higher if the step changes were assumed to be independent in space, because \emph{a-priori} assuming spatial clustering resulted in false step changes being identified close to real step changes.  This is an interesting result, as one may have naively assumed that for spatial data the step changes should be modelled spatially, which we have shown is not the case. Thus existing global smoothing models are sub-optimal for space time disease mapping in two respects, they smooth over such step changes leading to poorer estimation of disease risk, and they cannot identify such step changes which themselves provides etiological evidence about potential unmeasured risk factors.\\

\noindent Section 6 provided strong evidence of step changes in the unexplained component of risk for the England data. Additionally, better model fit with a smaller number of  effective parameters was observed compared to the global smoothing models, due to increased levels of smoothing in locations where step changes were absent. Thus non-adaptive smoothing models may overfit some data sets, by imposing too weak a spatial smoothing constraint due to the presence of step changes in risk. A striking association was found between the fitted risks and identified step changes between circulatory and respiratory disease, perhaps indicating the influence of the same unobserved risk factors.  Therefore in future work we will try and identify such unmeasured confounders, to see if they are indeed common to both diseases. A univariate approach to modelling was taken here as discussed in Section 2, but  the between disease correlation naturally suggests a bivariate spatio-temporal model as a future avenue of work. Such models have yet to be proposed in a spatio-temporal context, and would require the extension of multivariate spatial models such as multivariate conditional autoregressive (\citealp{gelfand2003proper}) models and shared component models (\citealp{knorr2001shared}). A final avenue for future work is to use the model in an ecological regression context, where the effect of an exposure on disease risk is of primary interest rather than the spatio-temporal pattern in disease risk. The efficacy of adaptive smoothing models in this context may be to reduce spatial confounding between the random effects and the covariates as suggested by \citet{clayton1993}, due to the reduction in the random effects variance (see Table 4),  and environmental factors such as air pollution would be a natural context for such work.

\section{Acknowledgements}
\noindent This work was funded by the Engineering and Physical Sciences Research Council (EPSRC), via grant EP/J017442/1, and the authors would like to thank the associate editor and a reviewer for their constructive comments that greatly improved the paper. 

\bibliographystyle{chicago}
\renewcommand{\bibname}{References}
\bibliography{references}

\end{document}

\footnotetext[1]{\it{Email address for correspondence: alastair.rushworth@strath.ac.uk}}
% 
% 

%%%%%%%%%%%%%%%%%%%%%%%%%%%%%%%%%%%%%%%%%%%%%%%%%%
%%%%%%%%%% INTRODUCTION 
%%%%%%%%%%%%%%%%%%%%%%%%%%%%%%%%%%%%%%%%%%%%%%%%%%

\section*{Introduction}
This supplementary material has the following sections. Section 1 provides an overview of computational details relating to the Markov chain Monte Carlo (McMC) algorithm used to implement the model proposed in this paper, while Section 2 presents example R code to apply the model proposed in the paper to simulated data.  Section 3 provides example code to implement the global smoothing models \textbf{(1)} and \textbf{(2)}.  Section 4 describes how to adjust the hyperparameters of the variance priors in order to assess prior sensitivity.
\newpage
\section{Fitting the localised spatio-temporal model}
One of the biggest computational overheads in the algorithm is updating $\mathbf{v}^{+}$ from it's full conditional distribution, which requires the evaluation of the log-density of $\bd{\phi}$ given by 
\begin{eqnarray*}
 \log(p(\bd{\phi}|\sigma^{2}, \epsilon, \mathbf{W})) & = & \frac{1}{2}\log|\sigma^{-2} \mathbf{Z}(\alpha) \otimes \mathbf{Q}(\mathbf{W}, \epsilon)| - \frac{1}{2\sigma^2} \bd{\phi}^{\top}[\mathbf{Z}(\alpha) \otimes  \mathbf{Q}(\mathbf{W}, \epsilon)] \bd{\phi}\\
 & = & \frac{1}{2}\sigma^{-2NT} + \frac{N}{2}\log| \mathbf{Z}(\alpha)| +  \frac{T}{2}\log|\mathbf{Q}(\mathbf{W}, \epsilon)|\\
&& - \frac{1}{2\sigma^{2}} \bd{\phi}^{\top}[\mathbf{Z}(\alpha) \otimes  \mathbf{Q}(\mathbf{W}, \epsilon)] \bd{\phi}.
 \end{eqnarray*}
 
\noindent Here $\mathbf{Z}(\alpha)$ is a $T \times T$ first order autoregressive temporal precision matrix that depends on the temporal autocorrelation parameter $\alpha$, and $\otimes$ denotes the Kronecker product.  As $\mathbf{W}$ depends on $\mathbf{v}^{+}$ many repeated evaluations of $|\mathbf{Q}(\mathbf{W}, \epsilon)|$ are required per iteration of the McMC algorithm.  To reduce the number of times that this operation is performed, a simple block proposal scheme is used, where blocks of $\mathbf{v}^{+}$ of size 10 are jointly accepted or rejected.  Matrix sparsity can be exploited to substantially decrease the time taken to evaluate $|\mathbf{Q}(\mathbf{W}, \epsilon)|$ and other operations involving $\mathbf{Q}(\mathbf{W}, \epsilon)$.  This is particularly important because although the matrix $\mathbf{Q}(\mathbf{W}, \epsilon)$ is complex and changes from one acceptance of  $\mathbf{v}^{+}$ to the next, it is highly sparse and therefore quadratic forms such as $\bd{\phi}^{\top}_j \mathbf{Q}(\mathbf{W}, \epsilon) \bd{\phi}_j$ and its determinants are available at much lower cost. Specifically, $|\mathbf{Q}(\mathbf{W}, \epsilon)|$ is available much more cheaply than by the $O(n^3)$ operation required to obtain its eigenvalues, simply by noting that 

\begin{eqnarray*}
|\mathbf{Q}(\mathbf{W}, \epsilon)| & = & 2\sum_{i=1}^N \log(L_{ii}),
 \end{eqnarray*}
 
\noindent where $\mathbf{L}$ is the Cholesky decomposition of $\mathbf{Q}(\mathbf{W}, \epsilon)$ such that $\mathbf{Q}(\mathbf{W}, \epsilon) = \mathbf{L}\mathbf{L}^{\top}$.  In addition, only a small proportion of the non-zero elements of $L$ need to be updated under a new block proposal, and therefore $\mathbf{L}$ does not need to be completely recalculated each time a  block of $\mathbf{v}^{+}$ is updated. 

\newpage
\section{Example R code to implement the adaptive model}

\subsection{Fit the adaptive model}

This section presents code to implement the model described in the paper.  The code uses R (version 3.2.2) and CRAN packages \texttt{Rcpp (0.12.1)}, \texttt{spam (1.0-1)}, \texttt{truncdist (1.0-1)} and \texttt{MCMCpack (1.3-3)}.  All of these necessary functions are imported on loading \texttt{CARBayesST} and so all that is required is to simply load the package
% \vspace{0.3cm}
% \begin{mdframed}[hidealllines=true,backgroundcolor=gray!20]
% {\setstretch{0.5}\color{blue}
% \texttt{library(CARBayesST)}}
% \end{mdframed}

\begin{knitrout}
\definecolor{shadecolor}{rgb}{0.969, 0.969, 0.969}\color{fgcolor}\begin{kframe}
\begin{alltt}
\hlkwd{library}\hlstd{(CARBayesST)}
\end{alltt}
\end{kframe}
\end{knitrout}

\noindent Various input data are required to fit the adaptive spatio-temporal model.  \newline\texttt{alldata.RData} contains the raw space time data in the form of a matrix, \texttt{alldata}, that is required to repeat the analysis of the English local authorities for circulatory and respiratory hospital admissions.  \texttt{alldata} contains several columns each of length $N \times T = 323 \times 10 = 3230$. \texttt{alldata\$observed.circulatory} and \texttt{alldata\$expected.circulatory}, are vectors of annual observed and expected circulatory hospital admissions for each of 323 English local health authority areas between 2001 and 2010.  The columns \texttt{alldata\$jobseekers}, \texttt{alldata\$urbanicity} and \texttt{alldata\$pm10} are the covariates for the proportion claiming jobseekers allowance, the urbanicity of each LHA and the mean PM$_{10}$ concentration.
% \vspace{0.3cm}
% \begin{mdframed}[hidealllines=true,backgroundcolor=gray!20]
% {\setstretch{1.0}\color{blue} \noindent \texttt{load("alldata.RData")}}
% \end{mdframed}

\begin{knitrout}
\definecolor{shadecolor}{rgb}{0.969, 0.969, 0.969}\color{fgcolor}\begin{kframe}
\begin{alltt}
\hlkwd{load}\hlstd{(}\hlstr{"alldata.RData"}\hlstd{)}
\end{alltt}
\end{kframe}
\end{knitrout}

\noindent \texttt{W.RData} contains the $323\times 323$ symmetric binary adjacency matrix \texttt{W} describing whether or not local authorities are adjacency to each other.

\begin{knitrout}
\definecolor{shadecolor}{rgb}{0.969, 0.969, 0.969}\color{fgcolor}\begin{kframe}
\begin{alltt}
\hlkwd{load}\hlstd{(}\hlstr{"W.RData"}\hlstd{)}
\end{alltt}
\end{kframe}
\end{knitrout}

% \vspace{0.3cm}
% \begin{mdframed}[hidealllines=true,backgroundcolor=gray!20]
% {\setstretch{1.0}\color{blue} \noindent \texttt{load("W.RData")}}
% \end{mdframed}

\noindent Using the above objects, the model is fitted using the function \texttt{ST.CARadaptive}.  The \texttt{formula} argument allows covariates and an offset term to be included in the model specification, following the conventions adopted in the \texttt{R} functions \texttt{lm} and \texttt{glm}.  \texttt{n.sample} defines the total length of the Markov chain, while \texttt{burnin} is the number of samples resulting from the Markov chain that are discarded as part of a `burn in' period.  The \texttt{thin} argument allows the user to specify a thinning schedule, so that only every $\texttt{thin}^{\textrm{th}}$ sample is stored, which is convenient if high correlation is observed between samples, or if storage space is limited.
% \vspace{0.3cm}
% \begin{mdframed}[hidealllines=true,backgroundcolor=gray!20]
% {\setstretch{0.2}\footnotesize\color{blue}
% \texttt{\# fit the model}
% \newline\texttt{model.circ <- ST.CARadaptive(}\\
% \indent \texttt{\hspace{1cm} formula = observed.circulatory $\sim$ jobseekers + pm10 + urbanicity + }\\
% \indent \texttt{\hspace{4cm} offset(log(expected.circulatory)),}\\
% \indent \texttt{\hspace{1cm} data ~~~~= alldata, }\\
% \indent \texttt{\hspace{1cm} family ~~= "poisson", }\\
% \indent \texttt{\hspace{1cm} W ~~~~~~~= W, }\\
% \indent \texttt{\hspace{1cm} n.sample = 100000, }\\
% \indent \texttt{\hspace{1cm} burnin ~~= 50000,}\\
% \indent \texttt{\hspace{1cm} thin ~~~~= 10)}
% }
% \end{mdframed}

\begin{knitrout}
\definecolor{shadecolor}{rgb}{0.969, 0.969, 0.969}\color{fgcolor}\begin{kframe}
\begin{alltt}
\hlcom{# fit the model}
\hlstd{model.circ} \hlkwb{<-} \hlkwd{ST.CARadaptive}\hlstd{(}
  \hlstd{observed.circulatory} \hlopt{\mytilde} \hlstd{jobseekers} \hlopt{+} \hlstd{pm10}
      \hlopt{+} \hlstd{urbanicity} \hlopt{+} \hlkwd{offset}\hlstd{(}\hlkwd{log}\hlstd{(expected.circulatory)),}
  \hlkwc{data} \hlstd{= alldata,}
  \hlkwc{family} \hlstd{=} \hlstr{"poisson"}\hlstd{,}
  \hlkwc{W} \hlstd{= W,}
  \hlkwc{n.sample} \hlstd{=} \hlnum{100000}\hlstd{,}
  \hlkwc{burnin} \hlstd{=} \hlnum{50000}\hlstd{,}
  \hlkwc{thin} \hlstd{=} \hlnum{10}\hlstd{)}
\end{alltt}
\end{kframe}
\end{knitrout}

\subsection{Plot the fitted model}

\noindent In order to plot step changes between adjacent local authorities, it is useful to have a list describing the line segments that compose the border shared by each pair of local authorities.  These are stored in the object \texttt{edge.list.RData} which contains the 2-column matrix, \texttt{edge.list} with $N_W$ rows describing the longitudes and latitudes of the vertices of the shared border between each pair of adjacent local authorities. 
% \vspace{0.3cm}
% \begin{mdframed}[hidealllines=true,backgroundcolor=gray!20]
% {\setstretch{1.0}\color{blue}\noindent\texttt{load("edge.list.RData")}}
% \end{mdframed}
\begin{knitrout}
\definecolor{shadecolor}{rgb}{0.969, 0.969, 0.969}\color{fgcolor}\begin{kframe}
\begin{alltt}
\hlkwd{load}\hlstd{(}\hlstr{"edge.list.RData"}\hlstd{)}
\end{alltt}
\end{kframe}
\end{knitrout}

\noindent Each border represented in \texttt{edge.list} corresponds to a single non-zero element in the upper triangle of \texttt{W}, and to link these two objects requires a look-up table.  This table is contained in  \texttt{edge.name.RData}, in the form of \texttt{edge.name} which is a character vector with the same length as \texttt{edge.list} ($N_W$).  The $i^{\textrm{th}}$ element of \texttt{edge.name} is a string of the form ``xxx.yyy'' that indicates that the $i^{\textrm{th}}$ border in \texttt{edge.list} is shared between areal units xxx and yyy and therefore corresponds to the adjacency element \texttt{W[xxx,yyy]}.
% \vspace{0.3cm}
% \begin{mdframed}[hidealllines=true,backgroundcolor=gray!20]
% {\setstretch{1.0}\color{blue}\texttt{load("edge.name.RData")}}
% \end{mdframed}
\begin{knitrout}
\definecolor{shadecolor}{rgb}{0.969, 0.969, 0.969}\color{fgcolor}\begin{kframe}
\begin{alltt}
\hlkwd{load}\hlstd{(}\hlstr{"edge.name.RData"}\hlstd{)}
\end{alltt}
\end{kframe}
\end{knitrout}

A final object, \texttt{poly\_list} is contained in \texttt{poly\_list.RData}, which is a list of length 323 describing the vertices outlining each of the polygons that define the perimeter of each local authority.  Each element of this list is a 2 column matrix of longitudes and latitudes, and retains the ordering of the local authorities used in \texttt{W} and the data.
% \vspace{0.3cm}
% \begin{mdframed}[hidealllines=true,backgroundcolor=gray!20]
% {\setstretch{1.0}\color{blue}\texttt{load("poly\_list.RData")}}
% \end{mdframed}

\begin{knitrout}
\definecolor{shadecolor}{rgb}{0.969, 0.969, 0.969}\color{fgcolor}\begin{kframe}
\begin{alltt}
\hlkwd{load}\hlstd{(}\hlstr{"poly_list.RData"}\hlstd{)}
\end{alltt}
\end{kframe}
\end{knitrout}

Extract the posterior median of the spatial random effects for the first year of study (2001) and create corresponding colour scale:
% \vspace{0.3cm}
% \begin{mdframed}[hidealllines=true,backgroundcolor=gray!20]
% {\footnotesize\setstretch{1.0}\color{blue}
% \texttt{exp.phi.hat <- exp(apply(model.circ\$samples\$phi, 2, median)[1:323])}\\
% \texttt{scale.brks ~<- seq(0, max(exp.phi.hat), length.out = 51)}\\
% \texttt{cols ~~~~~~~<- terrain.colors(n = 50)}\\
% \texttt{col.nums ~~~<- cut(exp.phi.hat, breaks = scale.brks, labels = FALSE)}\\
% \texttt{col.nums[is.na(col.nums)] <- 50}}
% \end{mdframed}

\begin{knitrout}
\definecolor{shadecolor}{rgb}{0.969, 0.969, 0.969}\color{fgcolor}\begin{kframe}
\begin{alltt}
\hlstd{exp.phi.hat} \hlkwb{<-} \hlkwd{exp}\hlstd{(}\hlkwd{apply}\hlstd{(model.circ}\hlopt{$}\hlstd{samples}\hlopt{$}\hlstd{phi,} \hlnum{2}\hlstd{, median)[}\hlnum{1}\hlopt{:}\hlnum{323}\hlstd{])}
\hlstd{scale.brks}  \hlkwb{<-} \hlkwd{seq}\hlstd{(}\hlnum{0}\hlstd{,} \hlkwd{max}\hlstd{(exp.phi.hat),} \hlkwc{length.out} \hlstd{=} \hlnum{51}\hlstd{)}
\hlstd{cols}        \hlkwb{<-} \hlkwd{terrain.colors}\hlstd{(}\hlkwc{n} \hlstd{=} \hlnum{50}\hlstd{)}
\hlstd{col.nums}    \hlkwb{<-} \hlkwd{cut}\hlstd{(exp.phi.hat,} \hlkwc{breaks} \hlstd{= scale.brks,} \hlkwc{labels} \hlstd{=} \hlnum{FALSE}\hlstd{)}
\hlstd{col.nums[}\hlkwd{is.na}\hlstd{(col.nums)]} \hlkwb{<-} \hlnum{50}
\end{alltt}
\end{kframe}
\end{knitrout}

Set up the plotting region and draw the areal units each with a colour corresponding to the associated posterior median of each spatial random effect:
% \vspace{0.3cm}
% \begin{mdframed}[hidealllines=true,backgroundcolor=gray!20]
% {\footnotesize\setstretch{1.0}\color{blue}
% \texttt{par(fig=c(0, 1, 0, 1),  oma = c(0, 0, 0, 5), mar = c(0, 0, 2, 5))}\\
% \texttt{plot(0, ylim = c(11075.8, 657599.5), xlim = c(134116.5, 655976.9), }\\
% \indent \texttt{ ~~~~~ type = "n", bty = "n", xaxt = "n", yaxt = "n")}\\
% \texttt{for(i in 1:length(poly\_list))\{ }\\
% \indent \texttt{ ~~~~~ polygon(poly\_list[[i]], col = cols[col.nums[i]], border = NA)}\\
% \}}
% \end{mdframed}

\begin{knitrout}
\definecolor{shadecolor}{rgb}{0.969, 0.969, 0.969}\color{fgcolor}\begin{kframe}
\begin{alltt}
\hlkwd{par}\hlstd{(}\hlkwc{fig}\hlstd{=}\hlkwd{c}\hlstd{(}\hlnum{0}\hlstd{,} \hlnum{1}\hlstd{,} \hlnum{0}\hlstd{,} \hlnum{1}\hlstd{),} \hlkwc{oma} \hlstd{=} \hlkwd{c}\hlstd{(}\hlnum{0}\hlstd{,} \hlnum{0}\hlstd{,} \hlnum{0}\hlstd{,} \hlnum{5}\hlstd{),} \hlkwc{mar} \hlstd{=} \hlkwd{c}\hlstd{(}\hlnum{0}\hlstd{,} \hlnum{0}\hlstd{,} \hlnum{2}\hlstd{,} \hlnum{5}\hlstd{))}
\hlkwd{plot}\hlstd{(}\hlnum{0}\hlstd{,} \hlkwc{ylim} \hlstd{=} \hlkwd{c}\hlstd{(}\hlnum{11075.8}\hlstd{,} \hlnum{657599.5}\hlstd{),} \hlkwc{xlim} \hlstd{=} \hlkwd{c}\hlstd{(}\hlnum{134116.5}\hlstd{,} \hlnum{655976.9}\hlstd{),}
\hlkwc{type} \hlstd{=} \hlstr{"n"}\hlstd{,} \hlkwc{bty} \hlstd{=} \hlstr{"n"}\hlstd{,} \hlkwc{xaxt} \hlstd{=} \hlstr{"n"}\hlstd{,} \hlkwc{yaxt} \hlstd{=} \hlstr{"n"}\hlstd{)}
\hlkwa{for}\hlstd{(i} \hlkwa{in} \hlnum{1}\hlopt{:}\hlkwd{length}\hlstd{(poly_list))\{}
 \hlkwd{polygon}\hlstd{(poly_list[[i]],} \hlkwc{col} \hlstd{= cols[col.nums[i]],} \hlkwc{border} \hlstd{=} \hlnum{NA}\hlstd{)}
 \hlstd{\}}
\end{alltt}
\end{kframe}
\end{knitrout}

Calculate summaries of the elements of $\mathbf{W}$ and plot step changes on the map:
% \vspace{0.3cm}
% \begin{mdframed}[hidealllines=true,backgroundcolor=gray!20]
% {\setstretch{1.0}\color{blue}
% \texttt{z       ~~~<- which(W > 0, arr.ind = T)}\\
% \texttt{z       ~~~<- z[which(z[,1] < z[,2]), ]}\\
% \texttt{inds    <- paste(z[,1], ".", z[,2], sep = "")}
% }
% \end{mdframed}

\begin{knitrout}
\definecolor{shadecolor}{rgb}{0.969, 0.969, 0.969}\color{fgcolor}\begin{kframe}
\begin{alltt}
\hlstd{z}    \hlkwb{<-} \hlkwd{which}\hlstd{(W} \hlopt{>} \hlnum{0}\hlstd{,} \hlkwc{arr.ind} \hlstd{= T)}
\hlstd{z}    \hlkwb{<-} \hlstd{z[}\hlkwd{which}\hlstd{(z[,}\hlnum{1}\hlstd{]} \hlopt{<} \hlstd{z[,}\hlnum{2}\hlstd{]), ]}
\hlstd{inds} \hlkwb{<-} \hlkwd{paste}\hlstd{(z[,}\hlnum{1}\hlstd{],} \hlstr{"."}\hlstd{, z[,}\hlnum{2}\hlstd{],} \hlkwc{sep} \hlstd{=} \hlstr{""}\hlstd{)}
\end{alltt}
\end{kframe}
\end{knitrout}

% \begin{mdframed}[hidealllines=true,backgroundcolor=gray!20]
% {\setstretch{1.0}\color{blue}
% \texttt{get\_prop\_thresh <- function(v, thresh)\{}\\
% \indent \texttt{ ~~~~~ as.numeric(!((sum(v < thresh)/length(v)) < 0.99))}\\
% \indent \texttt{\}}\\
% \indent \texttt{samples.w <- model.circ\$samples\$samples.w}\\
% \texttt{bdry <- apply(samples.w, 2, get\_prop\_thresh, thresh = 0.5)}\\
% \texttt{for(i in 1:length(edge.list))\{}\\
% \indent\texttt{ ~~~ if(bdry[i] ==  1)\{} \\
% \indent\texttt{ ~~~~~~edge.num <- which(edge.name == inds[i])}\\
% \indent\texttt{ ~~~~~~lines(edge.list[[edge.num]], lwd = 1, col = "white")}\\
% \indent\texttt{ ~~~\}}\\
% \indent\texttt{\}}
% }
% \end{mdframed}

\begin{knitrout}
\definecolor{shadecolor}{rgb}{0.969, 0.969, 0.969}\color{fgcolor}\begin{kframe}
\begin{alltt}
\hlstd{get_prop_thresh} \hlkwb{<-} \hlkwa{function}\hlstd{(}\hlkwc{v}\hlstd{,} \hlkwc{thresh}\hlstd{)\{}
  \hlkwd{as.numeric}\hlstd{(}\hlopt{!}\hlstd{((}\hlkwd{sum}\hlstd{(v} \hlopt{<} \hlstd{thresh)}\hlopt{/}\hlkwd{length}\hlstd{(v))} \hlopt{<} \hlnum{0.99}\hlstd{))}
\hlstd{\}}
\hlstd{samples.w} \hlkwb{<-} \hlstd{model.circ}\hlopt{$}\hlstd{samples}\hlopt{$}\hlstd{w}
\hlstd{bdry} \hlkwb{<-} \hlkwd{apply}\hlstd{(samples.w,} \hlnum{2}\hlstd{, get_prop_thresh,} \hlkwc{thresh} \hlstd{=} \hlnum{0.5}\hlstd{)}
\hlkwa{for}\hlstd{(i} \hlkwa{in} \hlnum{1}\hlopt{:}\hlkwd{length}\hlstd{(edge.list))\{}
  \hlkwa{if}\hlstd{(bdry[i]} \hlopt{==} \hlnum{1}\hlstd{)\{}
    \hlstd{edge.num} \hlkwb{<-} \hlkwd{which}\hlstd{(edge.name} \hlopt{==} \hlstd{inds[i])}
    \hlkwd{lines}\hlstd{(edge.list[[edge.num]],} \hlkwc{lwd} \hlstd{=} \hlnum{1}\hlstd{,} \hlkwc{col} \hlstd{=} \hlstr{"white"}\hlstd{)}
  \hlstd{\}}
\hlstd{\}}
\end{alltt}
\end{kframe}
\end{knitrout}

\vspace{0.3cm}
Finally, add a colour legend to the plot that identifies the risk level indicated by the colour shading in the local authority areas.
% \vspace{0.3cm}
% \begin{mdframed}[hidealllines=true,backgroundcolor=gray!20]
% {\footnotesize\setstretch{1.0}\color{blue}
% \texttt{par(fig = c(0.8,1,0.1,0.9), new = T)}\\
% \texttt{brks  <- seq(0, 200, length.out = 50)}\\
% \texttt{xvec  <- rep(0, 50)}\\
% \texttt{par(mar = c(0, 2.2, 0.5, 2.2), cex = 1)}\\
% \texttt{plot(scale.brks/500, scale.brks, type = "n", xaxs = "i",  }\\
% \indent     \texttt{ ~~~yaxs = "i", yaxt = "n", xlab = "", ylab = "", main = "", }\\
% \indent     \texttt{ ~~~bty = "n", xaxt = "n", xlim = c(-1, 1))}\\
% \texttt{tick.subset <- seq(1, 51, by = 5)}\\
% \texttt{axis(4, at = scale.brks[tick.subset], las = 1,}\\
% \indent     \texttt{ ~~~labels = round(c( scale.brks[tick.subset]), 2), ylog = T)}\\
% \texttt{rect(xvec, scale.brks[-length(scale.brks)], }\\
% \indent     \texttt{ ~~~xvec + 0.5, scale.brks[-1], col = cols, border = NA)}}
% \end{mdframed}   

\begin{knitrout}
\definecolor{shadecolor}{rgb}{0.969, 0.969, 0.969}\color{fgcolor}\begin{kframe}
\begin{alltt}
\hlkwd{par}\hlstd{(}\hlkwc{fig} \hlstd{=} \hlkwd{c}\hlstd{(}\hlnum{0.8}\hlstd{,}\hlnum{1}\hlstd{,}\hlnum{0.1}\hlstd{,}\hlnum{0.9}\hlstd{),} \hlkwc{new} \hlstd{= T)}
\hlstd{brks} \hlkwb{<-} \hlkwd{seq}\hlstd{(}\hlnum{0}\hlstd{,} \hlnum{200}\hlstd{,} \hlkwc{length.out} \hlstd{=} \hlnum{50}\hlstd{)}
\hlstd{xvec} \hlkwb{<-} \hlkwd{rep}\hlstd{(}\hlnum{0}\hlstd{,} \hlnum{50}\hlstd{)}
\hlkwd{par}\hlstd{(}\hlkwc{mar} \hlstd{=} \hlkwd{c}\hlstd{(}\hlnum{0}\hlstd{,} \hlnum{2.2}\hlstd{,} \hlnum{0.5}\hlstd{,} \hlnum{2.2}\hlstd{),} \hlkwc{cex} \hlstd{=} \hlnum{1}\hlstd{)}
\hlkwd{plot}\hlstd{(scale.brks}\hlopt{/}\hlnum{500}\hlstd{, scale.brks,} \hlkwc{type} \hlstd{=} \hlstr{"n"}\hlstd{,} \hlkwc{xaxs} \hlstd{=} \hlstr{"i"}\hlstd{,}
    \hlkwc{yaxs} \hlstd{=} \hlstr{"i"}\hlstd{,} \hlkwc{yaxt} \hlstd{=} \hlstr{"n"}\hlstd{,} \hlkwc{xlab} \hlstd{=} \hlstr{""}\hlstd{,} \hlkwc{ylab} \hlstd{=} \hlstr{""}\hlstd{,} \hlkwc{main} \hlstd{=} \hlstr{""}\hlstd{,}
    \hlkwc{bty} \hlstd{=} \hlstr{"n"}\hlstd{,} \hlkwc{xaxt} \hlstd{=} \hlstr{"n"}\hlstd{,} \hlkwc{xlim} \hlstd{=} \hlkwd{c}\hlstd{(}\hlopt{-}\hlnum{1}\hlstd{,} \hlnum{1}\hlstd{))}
\hlstd{tick.subset} \hlkwb{<-} \hlkwd{seq}\hlstd{(}\hlnum{1}\hlstd{,} \hlnum{51}\hlstd{,} \hlkwc{by} \hlstd{=} \hlnum{5}\hlstd{)}
\hlkwd{axis}\hlstd{(}\hlnum{4}\hlstd{,} \hlkwc{at} \hlstd{= scale.brks[tick.subset],} \hlkwc{las} \hlstd{=} \hlnum{1}\hlstd{,}
    \hlkwc{labels} \hlstd{=} \hlkwd{round}\hlstd{(}\hlkwd{c}\hlstd{( scale.brks[tick.subset]),} \hlnum{2}\hlstd{),} \hlkwc{ylog} \hlstd{= T)}
\hlkwd{rect}\hlstd{(xvec, scale.brks[}\hlopt{-}\hlkwd{length}\hlstd{(scale.brks)],}
    \hlstd{xvec} \hlopt{+} \hlnum{0.5}\hlstd{, scale.brks[}\hlopt{-}\hlnum{1}\hlstd{],} \hlkwc{col} \hlstd{= cols,} \hlkwc{border} \hlstd{=} \hlnum{NA}\hlstd{)}
\end{alltt}
\end{kframe}
\end{knitrout}

\newpage
\section{Example code to fit global smoothing models}

The non-adaptive spatio-temporal models that were compared in the modelling of the respiratory and circulatory data can also be fitted using functions from the \texttt{CARBayesST} \texttt{R} package.  Model \textbf{(1)} was the interaction model of \citet{knorr2000bayesianmodelling} and can be fitted using 
% \vspace{0.3cm}
% \begin{mdframed}[hidealllines=true,backgroundcolor=gray!20]
% {\setstretch{0.2}\footnotesize\color{blue}
% \texttt{\# fit the global Knorr-Held interaction model}
% \newline\texttt{model.circ <- ST.CARanova(}\\
% \indent \texttt{\hspace{1cm} formula = observed.circulatory $\sim$ jobseekers + pm10 + urbanicity + }\\
% \indent \texttt{\hspace{4cm} offset(log(expected.circulatory)),}\\
% \indent \texttt{\hspace{1cm} data ~~~~= alldata, }\\
% \indent \texttt{\hspace{1cm} family ~~= "poisson", }\\
% \indent \texttt{\hspace{1cm} W ~~~~~~~= W, }\\
% \indent \texttt{\hspace{1cm} n.sample = 100000, }\\
% \indent \texttt{\hspace{1cm} burnin ~~= 50000,}\\
% \indent \texttt{\hspace{1cm} thin ~~~~= 10)}
% }
% \end{mdframed}

\begin{knitrout}
\definecolor{shadecolor}{rgb}{0.969, 0.969, 0.969}\color{fgcolor}\begin{kframe}
\begin{alltt}
\hlcom{# fit the global Knorr-Held interaction model}
\hlstd{model.circ} \hlkwb{<-} \hlkwd{ST.CARanova}\hlstd{(}
  \hlstd{observed.circulatory} \hlopt{\mytilde} \hlstd{jobseekers} \hlopt{+} \hlstd{pm10}
    \hlopt{+} \hlstd{urbanicity} \hlopt{+} \hlkwd{offset}\hlstd{(}\hlkwd{log}\hlstd{(expected.circulatory)),}
  \hlkwc{data} \hlstd{= alldata,}
  \hlkwc{family} \hlstd{=} \hlstr{"poisson"}\hlstd{,}
  \hlkwc{W} \hlstd{= W,}
  \hlkwc{n.sample} \hlstd{=} \hlnum{100000}\hlstd{,}
  \hlkwc{burnin} \hlstd{=} \hlnum{50000}\hlstd{,}
  \hlkwc{thin} \hlstd{=} \hlnum{10}\hlstd{)}
\end{alltt}
\end{kframe}
\end{knitrout}

\vspace{0.3cm}
Model \textbf{(2)} was the interaction model of \citet{rushworth2014spatio} and can be fitted using 
% \vspace{0.3cm}
% \begin{mdframed}[hidealllines=true,backgroundcolor=gray!20]
% {\setstretch{0.2}\footnotesize\color{blue}
% \texttt{\# fit space-time Leroux model}
% \newline\texttt{model.circ <- ST.CARar(}\\
% \indent \texttt{\hspace{1cm} formula = observed.circulatory $\sim$ jobseekers + pm10 + urbanicity + }\\
% \indent \texttt{\hspace{4cm} offset(log(expected.circulatory)),}\\
% \indent \texttt{\hspace{1cm} data ~~~~= alldata, }\\
% \indent \texttt{\hspace{1cm} family ~~= "poisson", }\\
% \indent \texttt{\hspace{1cm} W ~~~~~~~= W, }\\
% \indent \texttt{\hspace{1cm} n.sample = 100000, }\\
% \indent \texttt{\hspace{1cm} burnin ~~= 50000,}\\
% \indent \texttt{\hspace{1cm} thin ~~~~= 10)}
% }
% \end{mdframed}

\begin{knitrout}
\definecolor{shadecolor}{rgb}{0.969, 0.969, 0.969}\color{fgcolor}\begin{kframe}
\begin{alltt}
\hlcom{# fit space-time Leroux model}
\hlstd{model.circ} \hlkwb{<-} \hlkwd{ST.CARar}\hlstd{(}
  \hlstd{observed.circulatory} \hlopt{\mytilde} \hlstd{jobseekers} \hlopt{+} \hlstd{pm10}
    \hlopt{+} \hlstd{urbanicity} \hlopt{+} \hlkwd{offset}\hlstd{(}\hlkwd{log}\hlstd{(expected.circulatory)),}
  \hlkwc{data} \hlstd{= alldata,}
  \hlkwc{family} \hlstd{=} \hlstr{"poisson"}\hlstd{,}
  \hlkwc{W} \hlstd{= W,}
  \hlkwc{n.sample} \hlstd{=} \hlnum{100000}\hlstd{,}
  \hlkwc{burnin} \hlstd{=} \hlnum{50000}\hlstd{,}
  \hlkwc{thin} \hlstd{=} \hlnum{10}\hlstd{)}
\end{alltt}
\end{kframe}
\end{knitrout}

\newpage
\section{Prior sensitivity in the adaptive model}

Sensitivity to different prior specifications can  be easily assessed by adjusting the hyperparameters for the relevant priors.  For example, in Section 1, the prior for $\beta_r$, is by default $\sim \mbox{N}(0, 1000)$, but this can be changed to a more or less informative value, for example $\sim \mbox{N}(0, 100)$ using the argument \texttt{prior.var.beta = ...}, for example:
% \vspace{0.3cm}
% \begin{mdframed}[hidealllines=true,backgroundcolor=gray!20]
% {\setstretch{0.2}\footnotesize\color{blue}
% \texttt{\# fit space-time adaptive model with N(0, 100) prior for fixed effect parameters}
% \newline\texttt{model.circ <- ST.CARadaptive(}\\
% \indent \texttt{\hspace{1cm} formula = observed.circulatory $\sim$ jobseekers + pm10 + urbanicity + }\\
% \indent \texttt{\hspace{4cm} offset(log(expected.circulatory)),}\\
% \indent \texttt{\hspace{1cm} data ~~~~= alldata, }\\
% \indent \texttt{\hspace{1cm} family ~~= "poisson", }\\
% \indent \texttt{\hspace{1cm} W ~~~~~~~= W, }\\
% \indent \texttt{\hspace{1cm} n.sample = 100000, burnin = 50000, thin = 10,}\\
% \indent \texttt{\hspace{1cm} prior.var.beta = rep(100, 4)}
% }
% \end{mdframed}

\begin{knitrout}
\definecolor{shadecolor}{rgb}{0.969, 0.969, 0.969}\color{fgcolor}\begin{kframe}
\begin{alltt}
\hlcom{# fit space-time adaptive model with N(0, 100) prior for fixed effect parameters}
\hlstd{model.circ} \hlkwb{<-} \hlkwd{ST.CARadaptive}\hlstd{(}
  \hlstd{observed.circulatory} \hlopt{\mytilde} \hlstd{jobseekers} \hlopt{+} \hlstd{pm10}
    \hlopt{+} \hlstd{urbanicity} \hlopt{+} \hlkwd{offset}\hlstd{(}\hlkwd{log}\hlstd{(expected.circulatory)),}
  \hlkwc{data} \hlstd{= alldata,}
  \hlkwc{family} \hlstd{=} \hlstr{"poisson"}\hlstd{,}
  \hlkwc{W} \hlstd{= W,}
  \hlkwc{n.sample} \hlstd{=} \hlnum{100000}\hlstd{,}
  \hlkwc{burnin} \hlstd{=} \hlnum{50000}\hlstd{,}
  \hlkwc{thin} \hlstd{=} \hlnum{10}\hlstd{,}
  \hlkwc{prior.var.beta} \hlstd{=} \hlkwd{rep}\hlstd{(}\hlnum{100}\hlstd{,} \hlnum{4}\hlstd{)}
\hlstd{)}
\end{alltt}
\end{kframe}
\end{knitrout}

Similarly, the default $IG(10^{-3},10^{-3})$ priors for $\zeta$ and $\tau$ can be changed to more informative $IG(a, b)$ using \texttt{prior.tau2 = c(a, b)}:
% \vspace{0.3cm}
% \begin{mdframed}[hidealllines=true,backgroundcolor=gray!20]
% {\setstretch{0.2}\footnotesize\color{blue}
% \texttt{\# fit space-time adaptive model with IG(a, b) prior for the variance parameters}
% \newline\texttt{model.circ <- ST.CARadaptive(}\\
% \indent \texttt{\hspace{1cm} formula = observed.circulatory $\sim$ jobseekers + pm10 + urbanicity + }\\
% \indent \texttt{\hspace{4cm} offset(log(expected.circulatory)),}\\
% \indent \texttt{\hspace{1cm} data ~~~~= alldata, }\\
% \indent \texttt{\hspace{1cm} family ~~= "poisson", }\\
% \indent \texttt{\hspace{1cm} W ~~~~~~~= W, }\\
% \indent \texttt{\hspace{1cm} n.sample = 100000, burnin = 50000, thin = 10,}\\
% \indent \texttt{\hspace{1cm} prior.tau2 = c(a, b)}\\
% )}
% \end{mdframed}

\begin{knitrout}
\definecolor{shadecolor}{rgb}{0.969, 0.969, 0.969}\color{fgcolor}\begin{kframe}
\begin{alltt}
\hlcom{# fit space-time adaptive model with IG(a, b) prior for the variance parameters}
\hlstd{model.circ} \hlkwb{<-} \hlkwd{ST.CARadaptive}\hlstd{(}
  \hlstd{observed.circulatory} \hlopt{\mytilde} \hlstd{jobseekers} \hlopt{+} \hlstd{pm10}
    \hlopt{+} \hlstd{urbanicity} \hlopt{+} \hlkwd{offset}\hlstd{(}\hlkwd{log}\hlstd{(expected.circulatory)),}
  \hlkwc{data} \hlstd{= alldata,}
  \hlkwc{family} \hlstd{=} \hlstr{"poisson"}\hlstd{,}
  \hlkwc{W} \hlstd{= W,}
  \hlkwc{n.sample} \hlstd{=} \hlnum{100000}\hlstd{,}
  \hlkwc{burnin} \hlstd{=} \hlnum{50000}\hlstd{,}
  \hlkwc{thin} \hlstd{=} \hlnum{10}\hlstd{,}
  \hlkwc{prior.tau2} \hlstd{=} \hlkwd{c}\hlstd{(a, b)}
\hlstd{)}
\end{alltt}
\end{kframe}
\end{knitrout}

\bibliographystyle{chicago}
\renewcommand{\bibname}{References}
\bibliography{references}